\documentclass[12pt, draftclsnofoot, onecolumn]{IEEEtran}
\usepackage{cite}

\usepackage{hyperref}
\usepackage{booktabs}
\usepackage{multirow}

\usepackage{verbatim}

\ifCLASSINFOpdf
   \usepackage[pdftex]{graphicx}
   \graphicspath{{../pdf/}{../jpeg/}}
   \DeclareGraphicsExtensions{.pdf,.jpeg,.png}
\else
   \usepackage[dvips]{graphicx}
   \graphicspath{{../eps/}}
   \DeclareGraphicsExtensions{.eps}
\fi
\usepackage{amsmath,amssymb}
\usepackage{algorithmic}
\makeatletter
\newcommand{\removelatexerror}{\let\@latex@error\@gobble}
\makeatother
\usepackage{algorithm}
\ifCLASSOPTIONcompsoc
  \usepackage[caption=false,font=normalsize,labelfont=sf,textfont=sf]{subfig}
\else
  \usepackage[caption=false,font=footnotesize]{subfig}
\fi
\begin{document}
%
\title{Speeding-up Symbol-Level Precoding Using Separable and Dual Optimizations}
%
%
%

\author{Junwen~Yang,
Ang~Li,~\IEEEmembership{Senior~Member,~IEEE,}
Xuewen~Liao,~\IEEEmembership{Member,~IEEE,}
and~Christos~Masouros,~\IEEEmembership{Senior~Member,~IEEE}
		
\thanks{Manuscript received XXXX; revised XXXX. \emph{(Corresponding author: Xuewen Liao.)}}

\thanks{J. Yang, A. Li are with the School of Information and Communications Engineering, Faculty of Electronic and Information Engineering, Xi'an Jiaotong University, Xi'an, Shaanxi 710049, China (e-mail: jwyang@stu.xjtu.edu.cn; ang.li.2020@xjtu.edu.cn).}

\thanks{X. Liao is with the School of Information and Communications Engineering, Faculty of Electronic and Information Engineering, Xi’an Jiaotong University, Xi’an, Shaanxi 710049, China, and also with the National Mobile Communications Research Laboratory, Southeast University, Nanjing 210096, China (e-mail: yeplos@mail.xjtu.edu.cn).}

\thanks{C. Masouros is with the Department of Electronic and Electrical Engineering, University College London, London WC1E 7JE, U.K. (e-mail: c.masouros@ucl.ac.uk).}}

\maketitle

\begin{abstract}
Symbol-level precoding (SLP) manipulates the transmitted signals to accurately exploit the multi-user interference (MUI) in the multi-user downlink. This enables that all the resultant interference contributes to correct detection, which is the so-called constructive interference (CI). Its performance superiority comes at the cost of solving a nonlinear optimization problem on a symbol-by-symbol basis, for which the resulting complexity becomes prohibitive in realistic wireless communication systems. In this paper, we investigate low-complexity SLP algorithms for both phase-shift keying (PSK) and quadrature amplitude modulation (QAM). Specifically, we first prove that the max-min SINR balancing (SB) SLP problem for PSK signaling is not separable, which is contrary to the power minimization (PM) SLP problem, and accordingly, existing decomposition methods are not applicable. Next, we establish an explicit duality between the PM-SLP and SB-SLP problems for PSK modulation. The proposed duality facilitates obtaining the solution to the SB-SLP given the solution to the PM-SLP without the need for one-dimension search, and vice versa. We then propose a closed-form power scaling algorithm to solve the SB-SLP via PM-SLP to take advantage of the separability of the PM-SLP. As for QAM modulation, we convert the PM-SLP problem into a separable equivalent optimization problem, and decompose the new problem into several simple parallel subproblems with closed-form solutions, leveraging the proximal Jacobian alternating direction method of multipliers (PJ-ADMM). We further prove that the proposed duality can be generalized to the multi-level modulation case, based on which a power scaling parallel inverse-free algorithm is also proposed to solve the SB-SLP for QAM signaling. Numerical results show that the proposed algorithms offer optimal performance with lower complexity than the state-of-the-art.
\end{abstract}


%
\IEEEpeerreviewmaketitle

\section{Introduction}
\label{secIntro}
\IEEEPARstart{I}{nterference}
is one of the major nuisances that deteriorate the performance of wireless communication systems \cite{5752452}. To achieve a promising performance for multi-user transmission in the downlink, precoding is recognized as an indispensable interference management technique at the transmitter side \cite{zheng2003diversity}. Conventional precoders in multi-antenna systems aim to suppress, mitigate, or eliminate interference because it distorts the desired signal just like noise. Taking the high-performance nonlinear precoders, for instance, dirty paper precoding (DPC) \cite{costa1983writing} and Tomlinson-Harashima precoding (THP) \cite{tomlinson1971new,harashima1972matched} compensate for the signal distortion induced by interference through pre-subtracting it successively, therefore the precoded signal is a nonlinear transformation on the data symbols. The vector perturbation (VP) precoding combines the regularization of channel inversion and the perturbation of transmit symbols. It jointly considers all the transmit symbols to generate an integer perturbation vector for data symbols using a sphere encoder, which differs from the DPC and THP that successively cancel the interference. Contrary to the nonlinear one, block-level precoding (BLP) generally uses only the channel state information (CSI) to calculate the precoding matrix, which is independent of the transmit data symbols. Zero-forcing (ZF) precoding is one of the typical linear precoders, which eliminates the multi-user interference via the inverse of the channel matrix \cite{caire2003achievable}. Such a simple operation will however augment the noise, thus limiting the performance. To this end, regularized ZF (RZF) precoding suppresses interference by introducing regularization to channel inversion \cite{peel2005vector}.

In addition to the above closed-form precoding, object-oriented linear precoding is another line of work, which designs the precoder involving specific quality of service (QoS) metrics and transmit power budgets. As it is generally modeled as a constrained optimization problem, also known as optimization-based precoding, e.g., the signal-to-interference-plus-noise (SINR)-constrained power minimization (PM) precoding \cite{visotsky1999optimum}, the power-constrained max-min SINR balancing (SB) precoding \cite{palomar2003joint,schubert2004solution,wiesel2005linear}, and the power-constrained weighted sum-rate (WSR) maximization precoding \cite{christensen2008weighted}. Early works focus on designing linear or block-level precoders through optimization, assuming independent and identically distributed (i.i.d.) data symbols. It has been revealed that the PM and SB problems are inverse problems \cite{wiesel2005linear}. Based on this inversion property, the SB problem can be solved by iteratively solving the PM problem for different SINR constraints along with a one-dimension bisection search \cite{wiesel2005linear}.

Only CSI is employed in the above closed-form linear precoding and optimization-based linear precoding. Nevertheless, information on the data symbols, which is also available at the transmitter, is not exploited for conventional block-level precoding. They ignore that once interference can be controlled instantaneously, it may be beneficial to signal detection \cite{masouros2015exploiting}. Therefore, instead of avoiding interference by leveraging the aforementioned conventional precoding schemes, recent works have proposed to exploit the known interference as a useful signal power based on the concept of constructive interference (CI) \cite{masouros2015exploiting}. Since data symbols vary among symbol slots, CI precoding is usually designed on a symbol-by-symbol basis, which is known as symbol-level precoding (SLP) \cite{7042789,li2020tutorial}. A seminal treatment of CI precoding was first proposed in a closed-form nonlinear precoding \cite{masouros2007novel}. The concept has been extended to optimization-based nonlinear precoding, attracting more and more attention because of the superior performance to its linear counterpart \cite{masouros2015exploiting,7042789,li2020tutorial}. 

The objective-oriented CI precoding concerning PM-SLP and SB-SLP problems was first proposed in \cite{7042789}, where all interference is strictly aligned with the data symbols, which followed the principle in \cite{masouros2010correlation}. Improvement has been made in \cite{masouros2015exploiting} by designing a more relaxed optimization for received symbols based on the proposed CI region. The above optimization-based CI-SLP works focus on only phase-shift keying (PSK) signaling. CI in quadrature amplitude modulation (QAM) constellation was first discussed in \cite{6619580}, and the first optimization-based CI-SLP concerning the PM problem for square multi-level modulation is studied in \cite{7942010}. The SB-SLP problem for multi-level modulation (or generic two-dimensional constellations) was investigated in \cite{8477154}.

Recent years have witnessed extensive endeavors to address low-complexity CI-SLP solutions. These include the efficient gradient projection algorithm to solve the Lagrangian dual problem of PM-SLP \cite{masouros2015exploiting}, closed-form suboptimal solutions for PM-SLP \cite{haqiqatnejad2018power,haqiqatnejad2019approximate}, derivations of the optimal precoding structure for SB-SLP with iterative algorithms \cite{li2018interference,li2020interference}, the CI-based BLP (CI-BLP) approach \cite{li2022practical,li2022block}, and the grouped SLP (G-SLP) approach \cite{xiao2022low}. More recently, the separability of the PM-SLP problem for PSK signaling has been revealed in \cite{yang2022low}, where parallel algorithms based on the alternating direction method of multipliers (ADMM) have been further proposed to solve the PM-SLP problem by exploiting its separable structure. However, the SB-SLP problem for PSK signaling still needs to be addressed, and whether it is separable is not yet clear. In addition, the existence or absence of separability in PM-SLP and SB-SLP problems for multi-level modulation remains unestablished. What's more, there is still a need to explore the parallel algorithms for multi-level modulation.

Motivated by the above observations, in this paper we address the SB-SLP problem for PSK signaling as well as the PM/SB-SLP problems for multi-level modulation. For clarity, the main contributions of this paper are summarized as follows:
\begin{enumerate}
\item  We first investigate the SB-SLP problem for PSK signaling. By rearranging the canonical problem formulation, we prove that this problem is not separable, hence it is not possible to decompose it into parallel subproblems. More importantly, we establish an explicit duality between the PM-SLP and SB-SLP problems for PSK signaling, which facilitates solving a pair of PM-SLP and SB-SLP problems simultaneously. This is a novel one-to-one mapping between the two problems, different from the existing inverse relation in \cite{7042789}. A one-step power scaling algorithm that solves the SB-SLP problem using the solution to the PM-SLP problem is developed, via which the parallel algorithms based on the separability are applicable to the SB-SLP problem.

\item  We then tackle the PM-SLP problem for multi-level modulation, for which the separability is analyzed. This problem is formulated as a nonlinear programming problem with equality and inequality CI constraints. Although separability is proven to exist for the PM-SLP problem with multi-level modulation, the parallel inverse-free SLP algorithms for PSK signaling are not directly feasible due to the inner constellation points. To obtain a low-complexity algorithm taking advantage of the separability, we introduce a slack variable converting the CI constraint points into equality. The feasible region of the slack variable is a polyhedral related to data symbols, which differs from the PM-SLP problem for PSK signaling considered in \cite{yang2022low}. The proximal Jacobian alternating direction method of multipliers (PJ-ADMM) is leveraged to solve the reformulated problem, arriving at a modified parallel inverse-free SLP (PIF-SLP) algorithm.

\item The SB-SLP problem for multi-level modulation is also studied. Similar to the PSK signaling case, we prove that this problem is not separable. The explicit duality of the two considered problems for multi-level modulation is further proven, therefore the proposed power scaling algorithm can also be applied to multi-level modulation. Based on the modified PIF-SLP algorithm and the power scaling algorithm, a power scaling PIF-SLP (SPIF-SLP) algorithm is proposed to solve the SB-SLP problem with multi-level modulation.
\end{enumerate}

Simulation results are conducted to validate our analysis of the separability and duality. They also demonstrate that the proposed parallelizable algorithms can greatly reduce the computational complexity of CI-SLP without sacrificing performance compared to existing works. Specifically, the most significant execution time reduction can be observed in the PM-SLP problem for multi-level modulation. Our algorithm is also shown to be competent for the SB-SLP problem for the challenging fully-loaded systems with multi-level modulation, although requiring more iterations than other scenarios.

The main novelty of this paper with respect to our previous work \cite{yang2022low} is the parallelizable methodology on the inseparable SB-SLP problem. Although the separability of the PM-SLP problem for PSK modulation has been considered in \cite{yang2022low} for the first time in the literature, and parallelizable algorithms have been therein proposed to reduce the complexity, the analysis and algorithms cannot directly be extended to the SB-SLP problems. This is because the PM-SLP and SB-SLP problems have different mathematical formulations and underlying rationales \cite{masouros2015exploiting,7042789,li2020tutorial}. To address this challenge, the methodology proposed herein resorts to dual and separable optimizations. We rigorously prove an explicit duality between the PM-SLP and weighted SB-SLP problems, which is used to solve the inseparable SB-SLP problem by the parallelizable method.

The remainder of this paper is organized as follows. Section \ref{secModel} introduces the system model and CI, as well as the canonical PM-SLP and SB-SLP problems for PSK signaling. Section \ref{secDuality} reformulates the two canonical problems in PSK modulation, establishes the proposed explicit duality between them, and further develops a closed-form power scaling algorithm for the SB-SLP problem. Section \ref{secQAM} addresses the PM-SLP and SB-SLP problems for QAM modulation, where we propose our PIF-SLP algorithm for the former and the SPIF-SLP algorithm for the latter. The explicit duality is also generalized to the QAM case.  Section \ref{secComplexity} provides the computational complexity analysis. Numerical results are presented in Section \ref{secResults}, and Section \ref{secConclusion} concludes the paper.

\textbf{Notation:} Scalars, vectors, and matrices, are represented by plain lowercase, boldface lowercase, and boldface capital letters, respectively. $(\cdot)^T$, $(\cdot)^H$, and $(\cdot)^{-1}$ denote transpose, conjugate transpose, and inverse operators, respectively. $\mathbb{C}^{M\times N}$ and $\mathbb{R}^{M\times N}$ denote the sets of $M\times N$ matrices with complex-valued and real-valued entries, respectively. $\left| \cdot \right|$ represents the absolute value of a real-valued scalar or the modulus of a complex-valued scalar. $\left\|\cdot\right\|$ denotes the Euclidean norm of a vector or spectral norm of a matrix. $\left\|\cdot\right\|_{\infty}$ represents the $\ell_{\infty}$-norm of a vector. $\Re \{\cdot\}$ and $\Im \{\cdot\}$ respectively denote the real part and imaginary part of a complex-valued input. $\succeq$ and $\unrhd$ denote the element-wise inequality and generalized inequality, respectively. $\mathbf{0}$, $\mathbf{1}$, and $\mathbf{I}$ represent respectively, the all-zeros vector, the all-ones vector, and the identity matrix with appropriate dimensions. $\oslash$ denotes the element-wise division. $diag\{\cdot\}$ returns a vector consisting of the main diagonal elements of a input matrix.

\section{System Model and Problem Formulation}
\label{secModel}
This section presents the system model and briefly reviews the concept of CI in the context of PSK signaling, followed by the PM-SLP and SB-SLP problem formulation with PSK modulation.
\subsection{System Model}
\label{subsecModel}
We consider a downlink multi-user multiple-input single-output (MU-MISO) system, where a base station (BS) equipped with $N_t$ antennas provides service for $K$ single-antenna users in the same time-frequency resource. The independent random data bits for each user are modulated to normalized data symbols. The data symbol vector $\tilde{\mathbf{s}}\triangleq [\tilde{s}_{1},\cdots , \tilde{s}_{K}]^{T}\in\mathbb{C}^{K}$ contains the overall $K$ data symbols in a symbol slot, which is mapped to the transmit signal $\tilde{\mathbf{x}}\triangleq [\tilde{x}_{1},\cdots , \tilde{x}_{N_{t}}]^{T}\in\mathbb{C}^{N_{t}}$ at the BS via SLP. The received signal of user $k$ in one symbol slot is expressed as
\begin{IEEEeqnarray}{rCl}
\tilde{y}_{k}&=&\tilde{\mathbf{h}}^T_{k}\tilde{\mathbf{x}}+\tilde{z}_{k} ,
\end{IEEEeqnarray}
where $\tilde{\mathbf{h}}_{k}\in\mathbb{C}^{N_t}$ denotes the quasi-static Rayleigh flat-fading channel vector between BS and user $k$, and $\tilde{z}_k\sim \mathcal{CN}(0, \sigma^2_k)$ is the complex-valued additive white Gaussian noise at user $k$. The channel matrix is denoted by $\tilde{\mathbf{H}}\triangleq[\tilde{\mathbf{h}}_{1},\cdots,\tilde{\mathbf{h}}_{K}]^T\in\mathbb{C}^{K\times N_{t}}$. To focus on the precoding design, perfect CSI is assumed.

\subsection{Constructive Interference}
\begin{figure}[!t]
\begin{minipage}[t]{0.5\linewidth}
\vspace{0pt}
\centering
\includegraphics{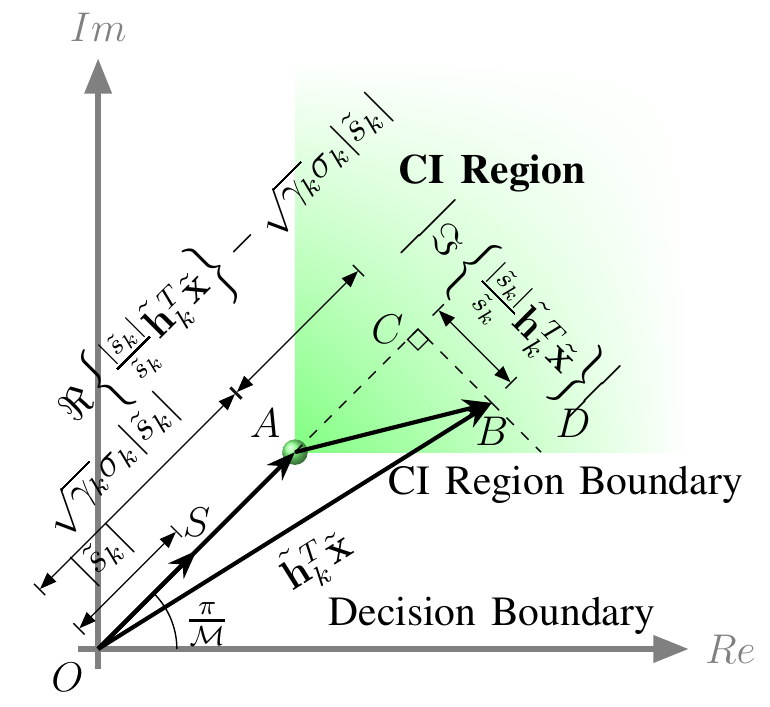}
\caption{Illustration of CI regions for a generic $\mathcal{M}$-PSK modulation.}
\label{fig_CI}
\end{minipage}
\hspace{0.1cm}
\begin{minipage}[t]{0.5\linewidth}
\vspace{0pt}
\centering
\includegraphics{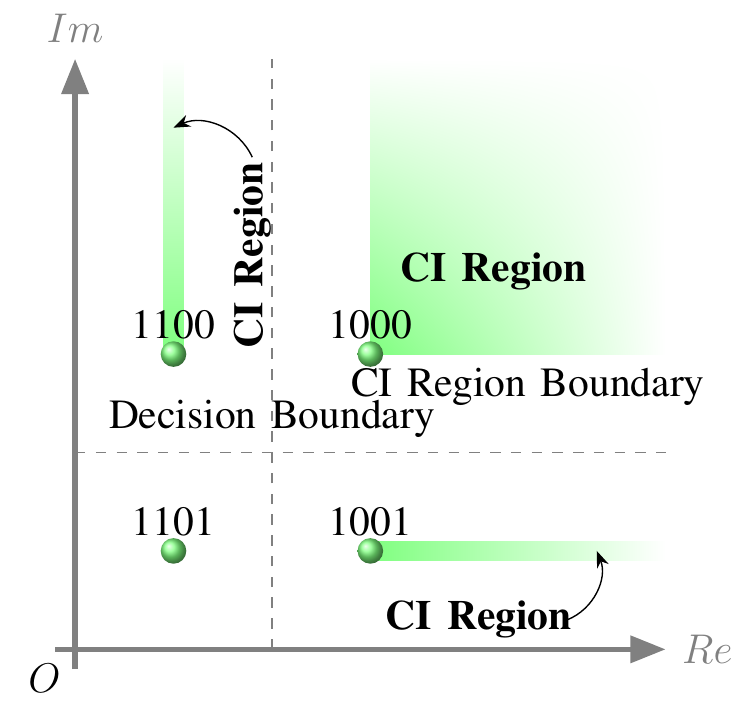}
\caption{Illustration of CI regions for a quarter of 16QAM modulation.}
\label{fig_CIQAM}
\end{minipage}
\end{figure}

To predict and further exploit the interference, CI precoding optimizes the transmit signal by judiciously utilizing CSI and data symbols, such that all the multi-user interference can add up constructively at each receiver side \cite{li2020tutorial}. As a consequence, the received instantaneous SINR at user $k$ is given as $\mathrm{SINR}_{k}=\frac{\vert\tilde{\mathbf{h}}^T_{k}\tilde{\mathbf{x}}\vert^2}{\sigma^2_k}$. Since all interference is exploited via CI precoding, the instantaneous SINR is equivalent to the conventional signal-to-noise ratio (SNR).

The optimization-based CI precoding attains CI leveraging the CI constraints. For the sake of illustration, the geometric interpretation of CI is shown in Fig. \ref{fig_CI}. Without loss of generality, denote the symbol of interest of user $k$ by $\tilde{s}_k$, which is an arbitrary constellation point drawn from a normalized $\mathcal{M}$-PSK constellation, corresponding to $\overrightarrow{OS}$. The received noiseless signal of user $k$ can be expressed as $\tilde{\mathbf{h}}^T_{k}\tilde{\mathbf{x}}$, which is denoted by $\overrightarrow{OB}$ in Fig. \ref{fig_CI}. For a given instantaneous SINR threshold $\gamma_k$ for user $k$, the nominal constellation point is equivalent to $\sqrt{\gamma_k}\sigma_k\tilde{s}_k$. We introduce $\overrightarrow{OA}$ as the nominal constellation point, which is also the only vertex of the interested CI region, where the CI region refers to a polyhedron bounded by hyperplanes parallel to decision boundaries or Voronoi edges of the constellation \cite{masouros2015exploiting,haqiqatnejad2018constructive}. The CI region associated with the nominal constellation point $\overrightarrow{OA}$ is depicted as the green-shaded area in Fig. \ref{fig_CI}. When $\overrightarrow{OB}$ is located in the depicted CI region, then the received signal is pushed away from decision boundaries, thus further into the correct decision region. In the meantime, the instantaneous SINR is guaranteed to be no less than the prescribed threshold $\gamma_k$. From Fig. \ref{fig_CI} we can observe that $\overrightarrow{OB}=\overrightarrow{OA}+\overrightarrow{AB}$. Furthermore, if $\overrightarrow{OB}$ is orthogonally decomposed along $\overrightarrow{OA}$, then we have $\overrightarrow{OB}=\overrightarrow{OC}+\overrightarrow{CB}$, where $\overrightarrow{OC}\perp\overrightarrow{CB}$. Consequently, one of the criteria that specifies the location of $\overrightarrow{OB}$ in the CI region is $\left|\overrightarrow{CD}\right|\geq \left|\overrightarrow{CB}\right|$, where $D$ denotes the intersection of $\overrightarrow{CB}$ and its nearest CI region boundary. The corresponding explicit mathematical formulation of CI constraints for $\mathcal{M}$-PSK signaling can be written as $
\Re\lbrace\hat{\mathbf{h}}^T_{k}\tilde{\mathbf{x}}\rbrace-\frac{\vert\Im\lbrace\hat{\mathbf{h}}^T_{k}\tilde{\mathbf{x}}\rbrace\vert}{\tan \frac{\pi}{\mathcal{M}}}\geq\sqrt{\gamma_k}\sigma_k,\, \forall k,
$
where $\hat{\mathbf{h}}^T_{k}\triangleq\frac{\tilde{\mathbf{h}}^T_{k}}{\tilde{s}_k}$, $\gamma_k$ denotes the pre-defined instantaneous SINR threshold for user $k$. It is worth noting that the SINR constraint for each user is already incorporated in the CI constraint.

\subsection{Problem Formulation}
\subsubsection{PM-SLP Problem}
\label{sec_PMformulation}
The PM-SLP problem aims to minimize the total transmit power subject to CI constraints. This optimization problem has the following mathematical form:
\begin{IEEEeqnarray}{rCl}
\label{eq_originalPM}
\begin{IEEEeqnarraybox}[][c]{rCl}
\min_{\tilde{\mathbf{x}}}&\,& \quad \left\|\tilde{\mathbf{x}}\right\|^2\\
\mathrm{s.t.}&\,& \Re\lbrace\hat{\mathbf{h}}^T_{k}\tilde{\mathbf{x}}\rbrace-\frac{\vert\Im\lbrace\hat{\mathbf{h}}^T_{k}\tilde{\mathbf{x}}\rbrace\vert}{\tan \frac{\pi}{\mathcal{M}}}\geq\sqrt{\gamma_k}\sigma_k,\, \forall k.
\end{IEEEeqnarraybox}
\end{IEEEeqnarray}
The above problem is convex and can be solved via off-the-shelf solvers. But most standard solvers, e.g., SeDuMi and SDPT3, are based on the high-complexity interior-point method (IPM). To alleviate the computational burden, recently a number of algorithms were proposed, e.g., the efficient algorithm based on gradient projection method \cite{masouros2015exploiting}, suboptimal closed-form solution \cite{haqiqatnejad2018power}, and improved suboptimal closed-form solution \cite{haqiqatnejad2019approximate}. 

More recently, we revealed the separability of the PM-SLP problem for PSK modulation and proposed a parallelizable and inversion-free CI-SLP precoding approach in our previous work \cite{yang2022low} based on the PJ-ADMM framework, which includes the PIF-SLP algorithm. In particular, by inspecting the problem and then rearranging the objective function and constraints into sums of multiple blocks, we revealed the separable structure of the PM-SLP problem for PSK modulation. This favorable structure facilitates decomposition, distributed, and even parallel computation methods. Therefore in \cite{yang2022low}, we converted the inequality CI constraints of (\ref{eq_originalPM}) into equality by introducing a nonnegative slack variable. Then we considered its augmented Lagrangian dual function, by which the ADMM framework can be used to decouple the optimization of the primal variable into multiple low-dimension subproblems. To parallelize each subproblem, we further added a proximal term to the augmented Lagrangian function. In addition, the second-order terms in the augmented Lagrangian function that incur matrix inversions to solutions can be eliminated by designing a proper proximal term. We also derived the closed-form solution to the slack variable. The above leads to the PIF-SLP algorithm for the PM-SLP problem for PSK modulation. More details please refer to \cite{yang2022low} and the references therein.

\subsubsection{SB-SLP Problem}
As mentioned in Section \ref{secIntro}, another typical SLP is the SB-SLP problem, which focuses on fairness in the system by maximizing the minimum instantaneous SINR over all users subject to a total transmit power constraint. This problem is formulated as
\begin{IEEEeqnarray}{rCl}
\label{eq_originalSB}
\begin{IEEEeqnarraybox}[][c]{rCl}
\max_{\tilde{\mathbf{x}}}\min_{k}&\,& \quad \frac{1}{\sqrt{\gamma_k}\sigma_k}\left\{\Re\lbrace\hat{\mathbf{h}}^T_{k}\tilde{\mathbf{x}}\rbrace-\frac{\Im\lbrace\hat{\mathbf{h}}^T_{k}\tilde{\mathbf{x}}\rbrace}{\tan \frac{\pi}{\mathcal{M}}},\Re\lbrace\hat{\mathbf{h}}^T_{k}\tilde{\mathbf{x}}\rbrace+\frac{\Im\lbrace\hat{\mathbf{h}}^T_{k}\tilde{\mathbf{x}}\rbrace}{\tan \frac{\pi}{\mathcal{M}}}\right\}\\
\mathrm{s.t.}&\,& \left\|\tilde{\mathbf{x}}\right\|^2\leq p,
\end{IEEEeqnarraybox}
\end{IEEEeqnarray}
where $p$ denotes the total transmit power budget, and, with a little abuse of notation, $\frac{1}{\sqrt{\gamma_k}}$ denotes the square root of the weight of $\mathrm{SINR}_k$ in the context of the SB-SLP problem.

The original max-min SB-SLP problem can be equivalently converted to a more tractable standard second-order cone programming (SOCP) problem \cite{masouros2015exploiting}, given by
\begin{IEEEeqnarray}{rCl}
\label{eq_originalSBSOCP}
\begin{IEEEeqnarraybox}[][c]{rCl}
\max_{\tilde{\mathbf{x}},\mu}&\,& \quad \mu\\
\mathrm{s.t.}&\,& \Re\lbrace\hat{\mathbf{h}}^T_{k}\tilde{\mathbf{x}}\rbrace-\frac{\vert\Im\lbrace\hat{\mathbf{h}}^T_{k}\tilde{\mathbf{x}}\rbrace\vert}{\tan \frac{\pi}{\mathcal{M}}}\geq \mu\sqrt{\gamma_k}\sigma_k,\, \forall k,\\
&\, &\left\|\tilde{\mathbf{x}}\right\|^2\leq p.
\end{IEEEeqnarraybox}
\end{IEEEeqnarray}
Similar to the PM-SLP problem, the problem above can be solved using standard solvers for convex optimization. The SB-SLP in SOCP form is more complex than the linearly constrained quadratic PM-SLP problem, although the PM-SLP problem is also SOCP. To solve it more efficiently, the derivation of the optimal structure is used to obtain its Lagrangian dual problem, which is shown to be quadratic over the probability simplex\cite{li2018interference}. Based on the solution structure analysis, an iterative closed-form scheme was proposed in \cite{li2018interference} for PSK signaling.

It is then natural to ask whether the SB-SLP problem is separable or not because this is essential for the employment of the PIF-SLP algorithm. To answer this question, Section \ref{secDuality} shows that separability does not exist in the SB-SLP problem for PSK modulation. By deriving an explicit duality between the two problems, Section \ref{secDuality} further proposes a closed-form one-step power scaling algorithm to solve the SB-SLP problem, provided that the solution to the PM-SLP problem is given.

\section{Proposed Closed-Form Power Scaling Algorithm for SB-SLP}
\label{secDuality}
This section reformulates the SB-SLP and PM-SLP problems for PSK modulation, then derives the one-to-one mapping between a pair of PM-SLP and SB-SLP problems for PSK modulation. Based on this a closed-form power scaling algorithm is proposed to solve the SB-SLP problem. Accordingly, the separable structure of the PM-SLP problem for PSK modulation can be employed in solving both the PM-SLP and SB-SLP problems.

\subsection{Problem Reformulation}
For notation simplicity, we adopt the equivalent real-valued notations. By using the complex-to-real transformation, the real-valued equivalent of (\ref{eq_originalSBSOCP}) can be written as
\begin{IEEEeqnarray}{rCl}
\label{eq_realSB}
\begin{IEEEeqnarraybox}[][c]{rCl}
\max_{\mathbf{x},\mu}&\,& \quad \mu\\
\mathrm{s.t.}&\,& \mathbf{T}\mathbf{S}_{k}\mathbf{H}_{k}\mathbf{x}\succeq\mu\sqrt{\gamma_k}\sigma_k\mathbf{1},\, \forall k,\\
&& \left\|\mathbf{x}\right\|^2\leq p,
\end{IEEEeqnarraybox}
\end{IEEEeqnarray}
where $\mathbf{x}\triangleq{\left[\begin{array}{c}\Re \lbrace {\tilde{\mathbf{x}}}\rbrace \\
\Im \lbrace {\tilde{\mathbf{x}}}\rbrace \end{array}\right]}\in\mathbb{R}^{2N_t},
\mathbf{T}\triangleq{\left[\begin{array}{cc}1 & -\frac{1}{\tan\frac{\pi}{\mathcal{M}}} \\
1 & \frac{1}{\tan\frac{\pi}{\mathcal{M}}} \end{array}\right]}\in\mathbb{R}^{2\times 2},
\mathbf{S}_{k}\triangleq{\left[\begin{IEEEeqnarraybox*}[][c]{,c/c,}
\Re \left\{\frac{1}{\tilde {s}_k}\right\} & -\Im \left\{\frac{1}{\tilde {s}_k}\right\} \\
\Im \left\{\frac{1}{\tilde {s}_k}\right\} & \Re \left\{\frac{1}{\tilde {s}_k}\right\} \end{IEEEeqnarraybox*}\right]}\in\mathbb{R}^{2\times 2},
\mathbf{H}_{k}\triangleq{\left[\begin{IEEEeqnarraybox*}[][c]{,c/c,}
\Re \lbrace {\tilde{\mathbf {h}}}^T_k\rbrace & -\Im \lbrace {\tilde{\mathbf{h}}}^T_k\rbrace \\ 
\Im \lbrace {\tilde{\mathbf {h}}}^T_k\rbrace & \Re \lbrace {\tilde{\mathbf {h}}}^T_k\rbrace \end{IEEEeqnarraybox*}\right]}\in\mathbb{R}^{2\times 2N_t}$. We further introduce $\bar{\mathbf{A}}_k\triangleq\mathbf{T}\mathbf{S}_{k}\mathbf{H}_{k}$, and $\mathbf{b}_k\triangleq\sqrt{\gamma_k}\sigma_k\mathbf{1}$. Accordingly, the CI constraints become $\bar{\mathbf{A}}_{k}\mathbf{x}\succeq\mu\mathbf{b}_k,\, \forall k$. A compact formulation can be attained by stacking the CI constraints over all the $K$ users, given by
\begin{IEEEeqnarray}{rCl}
\label{inequality constraint}
\mathbf{A}\mathbf{x}\succeq\mu\mathbf{b},
\end{IEEEeqnarray}
where $\mathbf{A}\triangleq\left[\bar{\mathbf{A}}^T_1,\cdots,\bar{\mathbf{A}}^T_K\right]^T\in\mathbb{R}^{2K\times 2N_t}$, $\mathbf{b}\triangleq\left[\mathbf{b}^T_1,\cdots,\mathbf{b}^T_K\right]^T\in\mathbb{R}^{2K}$. It can be seen that the left-hand side of (\ref{inequality constraint}) can be expressed as a linear combination of the columns of $\mathbf{A}$, i.e., $\sum_{i=1}^{2N_t}\mathbf{a}_{i}x_i$, where $\mathbf{a}_i$ is the $i$-th column of $\mathbf{A}$, $x_i$ is the $i$-th entry of $\mathbf{x}$. Subsequently, (\ref{eq_realSB}) can be rearranged as
\begin{IEEEeqnarray}{rCl}
\begin{IEEEeqnarraybox}[][c]{rCl}
\label{eq_separableSB}
\max_{\mathbf{x}_i,\mu} &\,& \quad \mu\\
\mathrm{s.t.} &\,& \sum_{i=1}^{N}\mathbf{A}_{i}\mathbf{x}_i\succeq\mu\mathbf{b},\\
&& \sum_{i=1}^{N}\left\|\mathbf{x}_i\right\|^2\leq p,
\end{IEEEeqnarraybox}
\end{IEEEeqnarray}
where $\mathbf{x}_i\in\mathbb{R}^{n_i}$ with $\sum^N_{i=1}n_i=2N_t$ and $\mathbf{A}_i\in\mathbb{R}^{2K\times n_i}$ are the $i$-th blocks of $\mathbf{x}$ and $\mathbf{A}$, respectively. $\mathbf{x}_i$ is composed of the adjacent and/or disadjacent elements of $\mathbf{x}$. Each column of $\mathbf{A}_i$ is uniquely taken from the columns of $\mathbf{A}$. Specifically, if the elements in $\mathbf{x}_i$ are taken from $\mathbf{x}$ continuously, we have $\mathbf{x}=\left[\mathbf{x}^T_1,\cdots,\mathbf{x}^T_N\right]^T$, $\mathbf{A}=\left[\mathbf{A}_1,\cdots,\mathbf{A}_N\right]$. On the other hand, if we want to group the disadjacent elements of $\mathbf{x}$ into one group, e.g., the real and imaginary parts of the same antenna, which can be expressed as $\mathbf{x}_i=\mathbf{E}^T_i\mathbf{x}$, $\mathbf{A}_i=\mathbf{A}\mathbf{E}_i$, where $\mathbf{E}_i\in\mathbb{R}^{2N_t\times n_i}$, and each column of $\left\{\mathbf{E}_i\right\}$ is uniquely picked from the columns of the $2N_t\times 2N_t$ identity matrix.

In accordance with the procedure formulating (\ref{eq_separableSB}), the real-valued equivalent of the PM-SLP problem (\ref{eq_originalPM}) can be rearranged as \cite{yang2022low}
\begin{IEEEeqnarray}{rCl}
\begin{IEEEeqnarraybox}[][c]{rCl}
\label{eq_separablePM}
\min_{\mathbf{x}_i} &\,& \quad \sum_{i=1}^{N}\left\|\mathbf{x}_i\right\|^2\\
\mathrm{s.t.} &\,& \sum_{i=1}^{N}\mathbf{A}_{i}\mathbf{x}_i\succeq\mathbf{b}.
\end{IEEEeqnarraybox}
\end{IEEEeqnarray}

The above formulation was first proposed in our previous work \cite{yang2022low}, where the separable structure of the PM-SLP problem for PSK modulation is proved. The structure was further utilized to decompose the original problem into multiple parallel subproblems by the proposed PIF-SLP algorithm.

Contrary to the separable PM-SLP problem for PSK modulation (\ref{eq_separablePM}), it is observed that the above SB-SLP problem (\ref{eq_separableSB}) is not separable because of the objective function $\mu$, which cannot be separated. Thus the PIF-SLP approach proposed in \cite{yang2022low} is not applicable to decompose the SB-SLP problem at first glance. Fortunately, we find an explicit relation inherent in the two problems, which indicates that once the optimal solution to the PM-SLP problem is obtained via the PIF-SLP \cite{yang2022low} or other algorithms, then finding the optimal solution to the SB-SLP problem is trivial, which is termed as the duality to be presented below.

\subsection{Duality Between the PM-SLP and SB-SLP for PSK Modulation}
For the conventional block-level interference suppression precoding, it is known that the PM problem and the SB problem are a pair of inverse problems \cite{wiesel2005linear,karipidis2008quality}. This relationship has been extended to CI-based SLP by \cite{7042789}, which proposes to solve the SB-SLP problem via iteratively solving its inverse PM-SLP problem along with a bisection search. Unlike the high-complexity one-dimension search scheme, recently, a novel duality between the conventional multicast PM and SB problems has been revealed \cite{7880721}, which explicitly determines the solution to the SB problem given the solution to the PM problem, and vice versa. Later in CI-based symbol error rate minimization precoding, a closed-form algorithm was designed to solve the detection-region-based noise uncertainty radius maximization problem under the precondition of the solved detection-region-based PM problem \cite{8374931}. In this subsection, we shall establish a novel duality between the PM-SLP and SB-SLP problems for PSK signaling.

Let $\mathbf{x}^{PM}$ and $p^{PM}\triangleq\|\mathbf{x}^{PM}\|^2$ denote the optimal solution and objective value of the PM-SLP problem for PSK modulation (\ref{eq_separablePM}). $\mathbf{x}^{SB}$ and $\mu^{SB}\triangleq \min\limits_{i}\frac{1}{\bar{b}_i}\bar{\mathbf{a}}^T_i \mathbf{x}^{SB}$ are the optimal counterparts for the SB-SLP problem in (\ref{eq_separableSB}), where $\bar{\mathbf{a}}_i$ denotes the transpose of the $i$-th row of $\mathbf{A}$, and $\bar{b}_i$ represents the $i$-th entry of $\mathbf{b}$.
\newtheorem{lemma}{\bf Lemma}
\begin{lemma}
\label{lemma_dualityvar}
The PM-SLP problem (\ref{eq_separablePM}) and the SB-SLP problem (\ref{eq_separableSB}) are inverse problems:
\begin{IEEEeqnarray}{rCl}
\label{eq_PMSBrela}
\mathbf{x}^{PM}\left(\alpha \mathbf{b}\right)&=&\mathbf{x}^{SB}\left(\mathbf{b},p^{PM}\left(\alpha \mathbf{b}\right)\right),
\end{IEEEeqnarray}
with $\alpha=\mu^{SB}\left(\mathbf{b},p^{PM}\left(\alpha \mathbf{b}\right)\right)$. Reciprocally,
\begin{IEEEeqnarray}{rCl}
\label{eq_SBPMrela}
\mathbf{x}^{SB}\left(\mathbf{b},p\right)&=&\mathbf{x}^{PM}\left(\mu^{SB}\left(\mathbf{b},p\right)\mathbf{b}\right),
\end{IEEEeqnarray}
with $p=p^{PM}\left(\mu^{SB}\left(\mathbf{b},p\right)\mathbf{b}\right)$.
\end{lemma}
\begin{IEEEproof}[\bf Proof]
Contradiction can be used to prove (\ref{eq_PMSBrela}). Assume that there exists an optimal solution $\mathbf{x}^{SB}\left(\mathbf{b},p^{PM}\left(\alpha \mathbf{b}\right)\right)$ and the corresponding optimal value $\mu^{SB}\left(\mathbf{b},p^{PM}\left(\alpha \mathbf{b}\right)\right)$ for the SB-SLP problem (\ref{eq_separableSB}) given parameters $\left(\mathbf{b},p^{PM}\left(\alpha \mathbf{b}\right)\right)$. Similarly, assume the optimal solution and the optimal value for the PM-SLP problem (\ref{eq_separablePM}) given $\alpha\mathbf{b}$ are $\mathbf{x}^{PM}\left(\alpha \mathbf{b}\right)$ and $p^{PM}\left(\alpha \mathbf{b}\right)$, respectively. By definition, $\mathbf{x}^{PM}\left(\alpha \mathbf{b}\right)$ is a feasible solution to the above SB-SLP problem, and the associated objective value is $\alpha$. If $\alpha>\mu^{SB}\left(\mathbf{b},p^{PM}\left(\alpha \mathbf{b}\right)\right)$, then this is a contradiction for the optimality of $\mu^{SB}\left(\mathbf{b},p^{PM}\left(\alpha \mathbf{b}\right)\right)$. Otherwise, if $\alpha<\mu^{SB}\left(\mathbf{b},p^{PM}\left(\alpha \mathbf{b}\right)\right)$, then $\mathbf{x}^{SB}\left(\mathbf{b},p^{PM}\left(\alpha \mathbf{b}\right)\right)$ is also a feasible solution to the PM-SLP problem (\ref{eq_separablePM}) given $\alpha\mathbf{b}$, for which all the CI constraints are over satisfied. Therefore, one can always find a $v\in(0,1)$ such that $v\mathbf{x}^{SB}\left(\mathbf{b},p^{PM}\left(\alpha \mathbf{b}\right)\right)$ meets all the CI constraints while providing a smaller objective value than $p^{PM}\left(\alpha \mathbf{b}\right)$. This is a contradiction for the optimality of $p^{PM}\left(\alpha \mathbf{b}\right)$. The above proves (\ref{eq_PMSBrela}) is true with $\alpha=\mu^{SB}\left(\mathbf{b},p^{PM}\left(\alpha \mathbf{b}\right)\right)$. The proof of (\ref{eq_SBPMrela}) is similar and is therefore omitted.
\end{IEEEproof}

\begin{lemma}
\label{lemma_scal}
Consider the PM-SLP problem (\ref{eq_separablePM}), for any $\alpha>0$, we have
\begin{IEEEeqnarray}{rCl}
\label{eq_PMvar}
\mathbf{x}^{PM}\left(\alpha \mathbf{b}\right)&=&\alpha\mathbf{x}^{PM}\left(\mathbf{b}\right),\\
\label{eq_PMobj}
p^{PM}\left(\alpha \mathbf{b}\right)&=&\alpha^2 p^{PM}\left(\mathbf{b}\right).
\end{IEEEeqnarray}
For the SB-SLP problem,
\begin{IEEEeqnarray}{rCl}
\label{eq_SBvar}
\mathbf{x}^{SB}\left(\mathbf{b},\alpha^2 p\right)&=&\alpha\mathbf{x}^{SB}\left(\mathbf{b},p\right),\\
\label{eq_SBobj}
\mu^{SB}\left(\mathbf{b},\alpha^2 p\right)&=&\alpha \mu^{SB}\left(\mathbf{b},p\right).
\end{IEEEeqnarray}
\end{lemma}
\begin{IEEEproof}[\bf Proof]
Let $\mathbf{x}=\frac{\dot{\mathbf{x}}}{\alpha}$, where $\alpha>0$, $\dot{\mathbf{x}}=\alpha\mathbf{x}$. Replacing $\mathbf{x}$ in (\ref{eq_separablePM}) yields
\begin{IEEEeqnarray}{rCl}
\begin{IEEEeqnarraybox}[][c]{rCl}
\label{eq_separablePMdot}
\min_{\dot{\mathbf{x}}_i} &\,& \quad \sum_{i=1}^{N}\left\|\dot{\mathbf{x}}_i\right\|^2\\
\mathrm{s.t.} &\,& \sum_{i=1}^{N}\mathbf{A}_{i}\dot{\mathbf{x}}_i\succeq\alpha\mathbf{b},
\end{IEEEeqnarraybox}
\end{IEEEeqnarray}
then (\ref{eq_PMvar}) and (\ref{eq_PMobj}) follow immediately.

By substituting $\mathbf{x}=\frac{\dot{\mathbf{x}}}{\alpha}$ into (\ref{eq_separableSB}), we similarly obtain 
\begin{IEEEeqnarray}{rCl}
\begin{IEEEeqnarraybox}[][c]{rCl}
\max_{\dot{\mathbf{x}}_i,\mu}&\,& \quad \alpha \mu\\
\mathrm{s.t.}&\,& \sum_{i=1}^{N}\mathbf{A}_{i}\dot{\mathbf{x}}_i\succeq \alpha \mu\mathbf{b},\\
&\, &\sum_{i=1}^{N}\left\|\dot{\mathbf{x}}_i\right\|^2\leq \alpha^2 p,
\end{IEEEeqnarraybox}
\end{IEEEeqnarray}
which induces (\ref{eq_SBvar}) and (\ref{eq_SBobj}).
\end{IEEEproof}

\newtheorem{theorem}{\bf Theorem}
\begin{theorem}[Duality for PSK Modulation]
\label{theorem_explicit}
Let $\mathbf{x}^{PM}$ and $p^{PM}\triangleq\|\mathbf{x}^{PM}\|^2$ denote the optimal solution and the optimal value of the PM-SLP problem (\ref{eq_separablePM}), respectively. Then the counterparts of the SB-SLP problem, $\mathbf{x}^{SB}$ and $\mu^{SB}$, are determined as
\begin{IEEEeqnarray}{rCl}
\label{eq_SBPMvar}
\mathbf{x}^{SB}\left(\mathbf{b},p\right)&=&\sqrt{\frac{p}{p^{PM}\left(\mathbf{b}\right)}}\mathbf{x}^{PM}\left(\mathbf{b}\right),\\
\label{eq_SBPMobj}
\mu^{SB}\left(\mathbf{b},p\right)&=&\sqrt{\frac{p}{p^{PM}\left(\mathbf{b}\right)}}.
\end{IEEEeqnarray}
and vice versa as
\begin{IEEEeqnarray}{rCl}
\label{eq_PMSBvar}
\mathbf{x}^{PM}\left(\mathbf{b}\right)&=&\frac{1}{\mu^{SB}\left(\mathbf{b},p\right)}\mathbf{x}^{SB}\left(\mathbf{b},p\right),\\
\label{eq_PMSBobj}
p^{PM}\left(\mathbf{b}\right)&=&\frac{p}{\left(\mu^{SB}\left(\mathbf{b},p\right)\right)^2}.
\end{IEEEeqnarray}
\end{theorem}
\begin{IEEEproof}[\bf Proof]
The optimal solution to the SB-SLP problem can be equivalently written as
\begin{IEEEeqnarray}{rCl}
\mathbf{x}^{SB}\left(\mathbf{b}, p\right)&=&\mathbf{x}^{SB}\left(\mathbf{b}, \frac{p}{p^{PM}\left(\mathbf{b}\right)}p^{PM}\left(\mathbf{b}\right)\right).
\label{eq_profSBSBvar}
\end{IEEEeqnarray}
By using (\ref{eq_PMobj}) to transfer the transmit power budget in (\ref{eq_profSBSBvar}), we have
\begin{IEEEeqnarray}{rCl}
\mathbf{x}^{SB}\left(\mathbf{b}, \frac{p}{p^{PM}\left(\mathbf{b}\right)}p^{PM}\left(\mathbf{b}\right)\right)=\mathbf{x}^{SB}\left(\mathbf{b}, p^{PM}\left(\sqrt{\frac{p}{p^{PM}\left(\mathbf{b}\right)}}\mathbf{b}\right)\right).
\label{eq_profSBSBvar2}
\end{IEEEeqnarray}
Combining (\ref{eq_profSBSBvar2}) with (\ref{eq_PMSBrela}) yields
\begin{IEEEeqnarray}{rCl}
\mathbf{x}^{SB}\left(\mathbf{b}, p^{PM}\left(\sqrt{\frac{p}{p^{PM}\left(\mathbf{b}\right)}}\mathbf{b}\right)\right)=\mathbf{x}^{PM}\left(\sqrt{\frac{p}{p^{PM}\left(\mathbf{b}\right)}}\mathbf{b}\right).\nonumber\\
\label{eq_profPMSBvar}
\end{IEEEeqnarray}
From (\ref{eq_PMvar}) we have
\begin{IEEEeqnarray}{rCl}
\mathbf{x}^{PM}\left(\sqrt{\frac{p}{p^{PM}\left(\mathbf{b}\right)}}\mathbf{b}\right)=\sqrt{\frac{p}{p^{PM}\left(\mathbf{b}\right)}}\mathbf{x}^{PM}\left(\mathbf{b}\right).
\label{eq_profPMvar}
\end{IEEEeqnarray}
Hence (\ref{eq_SBPMvar}) is true.

We then use Lemma \ref{lemma_dualityvar} and Lemma \ref{lemma_scal} to prove (\ref{eq_SBPMobj}) and (\ref{eq_PMSBobj}). It is shown in Lemma \ref{lemma_dualityvar} that
\begin{IEEEeqnarray}{rCl}
p=p^{PM}\left(\mu^{SB}\left(\mathbf{b},p\right)\mathbf{b}\right).
\end{IEEEeqnarray}
Using (\ref{eq_PMobj}), the above equality yields
\begin{IEEEeqnarray}{rCl}
p^{PM}\left(\mu^{SB}\left(\mathbf{b},p\right)\mathbf{b}\right)=\left(\mu^{SB}\left(\mathbf{b},p\right)\right)^2 p^{PM}\left(\mathbf{b}\right).
\end{IEEEeqnarray}
Thus (\ref{eq_SBPMobj}) and (\ref{eq_PMSBobj}) follow immediately.

The proof of (\ref{eq_PMSBvar}) is similar to that of (\ref{eq_SBPMvar}). For brevity, we give an abbreviated proof below:
\begin{IEEEeqnarray}{rCl}
\mathbf{x}^{PM}\left(\mathbf{b}\right)&=&\mathbf{x}^{PM}\left(\frac{1}{\mu^{SB}\left(\mathbf{b},p\right)}t^{SB}\left(\mathbf{b},p\right)\mathbf{b}\right)
\overset{(\ref{eq_SBobj})}{=}\mathbf{x}^{PM}\left(\mu^{SB}\left(\mathbf{b},\frac{1}{\left(\mu^{SB}\left(\mathbf{b},p\right)\right)^2}p\right)\mathbf{b}\right)\nonumber\\
&\overset{(\ref{eq_SBPMrela})}{=}&\mathbf{x}^{SB}\left(\mathbf{b},\frac{1}{\left(\mu^{SB}\left(\mathbf{b},p\right)\right)^2}p\right)
\overset{(\ref{eq_SBvar})}{=}\frac{1}{\mu^{SB}\left(\mathbf{b},p\right)}\mathbf{x}^{SB}\left(\mathbf{b},p\right).
\end{IEEEeqnarray}
\end{IEEEproof}

\newtheorem{corollary}{\bf Corollary}
\begin{corollary}
\label{corollary_powscal}
The SB-SLP problem and the PM-SLP problem can be solved simultaneously. In particular, the solution to the SB-SLP problem (\ref{eq_separableSB}) can be obtained by first solving the PM-SLP problem (\ref{eq_separablePM}) and then scaling the transmit power to satisfy the power budget of the SB-SLP problem, and vice versa.
\end{corollary}
\subsection{Power Scaling Algorithm}
According to Corollary \ref{corollary_powscal}, the SB-SLP problem for PSK modulation can be solved by a simple one-step power scaling algorithm, provided that the solution to the PM-SLP problem is available. Specifically, we solve (\ref{eq_separablePM}) by the PIF-SLP \cite{yang2022low} or other algorithms and obtain $\mathbf{x}^{PM}(\mathbf{b})$ as well as $p^{PM}(\mathbf{b})$, then compute the solution to SB-SLP by (\ref{eq_SBPMvar}), which is termed the power scaling algorithm for SLP. We point out that although the parallel algorithm cannot directly be applied to the SB-SLP problem due to the lack of separability, a SPIF-SLP algorithm can be designed to solve the SB-SLP problem with the aid of the closed-form power scaling algorithm, which consists of two steps. In the first step, we obtain the parallelizable solution to the PM-SLP problem via the PIF-SLP algorithm proposed in \cite{yang2022low}, which is outlined in Section \ref{sec_PMformulation}. Whileas in the second step, we use the closed-form power scaling algorithm to acquire the solution to the SB-SLP problem. By applying the PIF-SLP algorithm along with the closed-form power scaling algorithm, the separability of the PM-SLP problem can be utilized to attain a low-complexity and parallelizable solution to the SB-SLP problem. 
	
\section{Proposed Parallelizable CI-SLP Algorithms for QAM Modulation}
\label{secQAM}
In this section, we address the PM-SLP and SB-SLP problems for QAM modulation. Other multi-level modulations such as amplitude phase shift keying (APSK) can be analyzed with a similar methodology.

To begin with, the first quadrant of a 16QAM constellation is depicted in Fig. \ref{fig_CIQAM} as an example, from which we can observe that only the inner constellation point `1101' has a closed or fully-bounded decision region, and the other three constellation points have open decision regions bounded by either two or three decision boundaries. The three darker-shaded areas associated with `1001', `1100', and `1000' are CI regions. One of the main differences between multi-level and constant-envelope modulations is whether the constellation point has closed decision regions or not. From Fig. \ref{fig_CIQAM} we can conclude that only the data symbols corresponding to open decision boundaries have degrees of freedom to exploit CI. Specifically, the open decision regions bounded by two and three decision boundaries have two and one dimensions to exploit CI, respectively. For the inner constellation points, we cannot push them away from one decision boundary while preserving the distance to another decision boundary. Therefore, as shown in Fig. \ref{fig_CIQAM}, the left-edge constellation point `1001' has its real part to exploit CI, and the upper-edge constellation point `1100' has its imaginary part to exploit CI. The corresponding CI regions are two rays. On the other hand, the vertex constellation point `1000' has both real and imaginary parts to exploit CI. Consequently, its CI region is a two-dimensional convex polyhedron similar to PSK modulation.

The mathematical formulation of CI constraints such that the noiseless received signal lies in the CI region and meets the instantaneous SINR threshold $\gamma_k$ for QAM signaling can be written as
\begin{IEEEeqnarray}{lCr}
sign\left\{
\Re\left\{\tilde{s}_k\right\}\right\}\Re\left\{\tilde{\mathbf{h}}^T_k\tilde{\mathbf{x}}\right\}&\unrhd & sign\left\{\Re\left\{\tilde{s}_k\right\}\right\}\sqrt{\gamma_k}\sigma_k\Re\left\{\tilde{s}_k\right\},\forall k,
\IEEEeqnarraynumspace\\
sign\left\{
\Im\left\{\tilde{s}_k\right\}\right\}\Im\left\{\tilde{\mathbf{h}}^T_k\tilde{\mathbf{x}}\right\}&\unrhd & sign\left\{\Im\left\{\tilde{s}_k\right\}\right\}\sqrt{\gamma_k}\sigma_k\Im\left\{\tilde{s}_k\right\},\forall k,
\IEEEeqnarraynumspace
\end{IEEEeqnarray}
where $\unrhd$ represents the generalized inequality symbol, i.e., $\unrhd$ equals to $\geq$ or $=$, depending on whether CI can be exploited or not. By introducing $\hat{\mathbf{h}}^T_{k}\triangleq\frac{\tilde{\mathbf{h}}^T_{k}}{\tilde{s}_k}$, the above original CI constraints can be rearranged as
\begin{IEEEeqnarray}{lCr}
\Re\left\{\hat{\mathbf{h}}^T_k\tilde{\mathbf{x}}\right\}&\unrhd & \sqrt{\gamma_k}\sigma_k, \forall k,\\
\Im\left\{\hat{\mathbf{h}}^T_k\tilde{\mathbf{x}}\right\}&\unrhd & \sqrt{\gamma_k}\sigma_k, \forall k.
\end{IEEEeqnarray}

\subsection{Problem Formulation}
\subsubsection{PM-SLP Problem}
The PM-SLP problem for QAM modulation that minimizes the total transmit power subject to CI constraints can be formulated as
\begin{IEEEeqnarray}{rCl}
\label{eq_originalPMQAM}
\begin{IEEEeqnarraybox}[][c]{rCl}
\min_{\tilde{\mathbf{x}}}&\,& \quad \left\|\tilde{\mathbf{x}}\right\|^2\\
\mathrm{s.t.}&\,& \Re\left\{\hat{\mathbf{h}}^T_k\tilde{\mathbf{x}}\right\}\unrhd  \sqrt{\gamma_k}\sigma_k, \forall k,\\
& &\Im\left\{\hat{\mathbf{h}}^T_k\tilde{\mathbf{x}}\right\}\unrhd  \sqrt{\gamma_k}\sigma_k, \forall k.
\end{IEEEeqnarraybox}
\end{IEEEeqnarray}
Although the efficient algorithm based on the gradient projection method proposed in \cite{masouros2015exploiting} was designed only for PSK modulation, it can be used to solve the above problem with proper modification. The suboptimal closed-form solution proposed in \cite{haqiqatnejad2018power} and the improved suboptimal closed-form solution proposed in \cite{haqiqatnejad2019approximate} can also be employed to obtain suboptimal solutions to the above problem.

\subsubsection{SB-SLP Problem}
The SB-SLP problem for QAM modulation aims to maximize the minimum instantaneous SINR in CI regions subject to a total transmit power constraint. Like in the PSK modulation case, this problem can also be rewritten in a SOCP form given by
\begin{IEEEeqnarray}{rCl}
\label{eq_originalSBQAM}
\begin{IEEEeqnarraybox}[][c]{rCl}
\max_{\tilde{\mathbf{x}},\mu}&\,& \quad \mu\\
\mathrm{s.t.}&\,& \frac{1}{\sqrt{\gamma_k}}\Re\left\{\hat{\mathbf{h}}^T_k\tilde{\mathbf{x}}\right\}\unrhd  \mu\sigma_k, \forall k,\\
& &\frac{1}{\sqrt{\gamma_k}}\Im\left\{\hat{\mathbf{h}}^T_k\tilde{\mathbf{x}}\right\}\unrhd  \mu\sigma_k, \forall k,\\
& &\left\|\tilde{\mathbf{x}}\right\|^2\leq p,
\end{IEEEeqnarraybox}
\end{IEEEeqnarray}
where $p$ denotes the total transmit power budget, $\frac{1}{\sqrt{\gamma_k}}$ denotes the square-root of the weight of $\mathrm{SINR}_k$. Following the iterative algorithm for the SB-SLP problem for PSK modulation \cite{li2018interference}, a modified iterative algorithm, as well as a suboptimal closed-form solution, were subsequently developed for the above problem\cite{li2020interference}.

\subsection{Separability of the PM-SLP for QAM Modulation}
The real-valued equivalent of (\ref{eq_originalPMQAM}) is given by
\begin{IEEEeqnarray}{rCl}
\label{eq_realPMQAM}
\begin{IEEEeqnarraybox}[][c]{rCl}
\min_{\mathbf{x}}&\,& \quad \left\|\mathbf{x}\right\|^2\\
\mathrm{s.t.}&\,& \mathbf{S}_{k}\mathbf{H}_{k}\mathbf{x}\unrhd\sqrt{\gamma_k}\sigma_k\mathbf{1},\, \forall k,
\end{IEEEeqnarraybox}
\end{IEEEeqnarray}
where $\mathbf{x}$, $\mathbf{S}_{k}$, and $\mathbf{H}_{k}$ have the same values as those in the PSK case. It can be seen that the above PM-SLP problem formulation for QAM modulation is equivalent to its counterpart for QPSK modulation, except for the generalized inequality symbol. This is because the four vertex constellation points in a square QAM constellation can be viewed as a QPSK constellation. The above problem can be rearranged to a separable formulation, given by
\begin{IEEEeqnarray}{rCl}
\begin{IEEEeqnarraybox}[][c]{rCl}
\label{eq_separablePMQAM}
\min_{\mathbf{x}_i} &\,& \quad \sum_{i=1}^{N}\left\|\mathbf{x}_i\right\|^2\\
\mathrm{s.t.} &\,& \sum_{i=1}^{N}\mathbf{A}_{i}\mathbf{x}_i\unrhd\mathbf{b},
\end{IEEEeqnarraybox}
\end{IEEEeqnarray}
where the notations directly inherent from the PSK case, except for $\mathbf{A}\triangleq\left[\bar{\mathbf{A}}^T_1,\cdots,\bar{\mathbf{A}}^T_K\right]^T\in\mathbb{R}^{2K\times 2N_t}$, $\bar{\mathbf{A}}_k\triangleq\mathbf{S}_{k}\mathbf{H}_{k}$. Accordingly, the optimization variable $\mathbf{x}$ is split into $N$ separate subvectors. In addition, the objective function and constraints of the PM-SLP problem for QAM modulation can also be written as summations of $N$ individual blocks, each of which only associates with a subvector of $\mathbf{x}$. This indicates that the problem is separable, similar to the PSK case. Decomposition methods are therefore applicable to partition the problem into smaller separate subproblems, each of which can be updated in a sequential or parallel, centralized or decentralized manner.

\subsection{Parallel Inverse-Free Algorithm for PM-SLP for QAM Modulation}
This subsection develops a PIF-SLP algorithm for the PM-SLP problem for QAM modulation taking advantage of its separability presented in the previous subsection. Although sharing the same name with our previous work in \cite{yang2022low} because they have the same parallelizable and inversion-free properties, this algorithm is different from the previous one. The reason lies in that we have to tackle both the inequality and equality constraints corresponding to the outer and inner constellation points for QAM modulation, which leads to a different feasible region for the Lagrangian multiplier compared to the PSK case.

To start with, we reformulate (\ref{eq_separablePMQAM}) by introducing a slack variable vector $\mathbf{c}$ to convert the original generalized inequality constraints into corresponding equality constraints as follows:
\begin{IEEEeqnarray}{rCl}
\begin{IEEEeqnarraybox}[][c]{rCl}
\min_{\mathbf{x}_i,\mathbf{c}} &\,& \quad \sum_{i=1}^{N}\left\|\mathbf{x}_i\right\|^2\\
\mathrm{s.t.}&\,& \sum_{i=1}^{N}\mathbf{A}_{i}\mathbf{x}_i=\mathbf{b}+\mathbf{c},\\
&&\mathbf{c}\in\mathcal{C},
\end{IEEEeqnarraybox}
\end{IEEEeqnarray}
where $\mathcal{C}\triangleq\left\{\mathbf{c}\big\vert c_i\geq0,\forall i\in\mathcal{W}; c_j=0,\forall j\neq i\right\}\subseteq\mathbb{R}^{2K}$ is the feasible region of $\mathbf{c}$, where $\mathcal{W}\triangleq\left\{i\big\vert|s_i|=\|\mathbf{w}\|_\infty\right\}$, $\mathbf{s}\triangleq\left[\Re\left\{\tilde{s}_1\right\},\Im\left\{\tilde{s}_1\right\},\cdots,\Re\left\{\tilde{s}_K\right\},\Im\left\{\tilde{s}_K\right\}\right]^T$, $c_i$ and $s_i$ are the $i$-th entries of $\mathbf{c}$ and $\mathbf{s}$, respectively. Assume $\mathbf{w}\triangleq\left[\Re\left\{\tilde{w}_1\right\},\Im\left\{\tilde{w}_1\right\},\cdots,\Re\left\{\tilde{w}_{\mathcal{M}}\right\},\Im\left\{\tilde{w}_{\mathcal{M}}\right\}\right]^T$ is composed of the real and imaginary parts of all the constellation points of a square $\mathcal{M}$-QAM constellation. The feasible constraint of the slack variable $\mathbf{c}$ can be further incorporated into the objective function:
\begin{IEEEeqnarray}{rCl}
\begin{IEEEeqnarraybox}[][c]{rCl}
\label{eq_slackSepara}
\min_{\mathbf{x}_i,\mathbf{c}}&\,& \quad \sum_{i=1}^{N}\left\|\mathbf{x}_i\right\|^2+\mathcal{I}_{\mathcal{C}}\left(\mathbf{c}\right)\\
\mathrm{s.t.}&\,& -\sum_{i=1}^{N}\mathbf{A}_{i}\mathbf{x}_i+\mathbf{b}+\mathbf{c}=\mathbf{0},
\end{IEEEeqnarraybox}
\end{IEEEeqnarray}
where $\mathcal{I}_{\mathcal{C}}\left(\mathbf{c}\right)$ is the indicator function of $\mathcal{C}$ given by
$
\mathcal{I}_{\mathcal{C}}\left(\mathbf{c}\right)=
\begin{cases}
0,&\text{if } \mathbf{c}\in\mathcal{C},\\
+\infty, &\text{otherwise.}
\end{cases}
$

The augmented Lagrangian function of (\ref{eq_slackSepara}) is given by 
\begin{IEEEeqnarray}{rCl}
\label{eq_slackAugmentedLagrangian}
\mathcal{L}_\rho\left(\mathbf{x},\mathbf{c},\boldsymbol{\lambda}\right)&=&\sum_{i=1}^{N}\left\|\mathbf{x}_i\right\|^2+I_{\mathcal{C}}\left(\mathbf{c}\right)+\boldsymbol{\lambda}^T\left(-\sum_{i=1}^{N}\mathbf{A}_{i}\mathbf{x}_i+\mathbf{b}+\mathbf{c}\right)+\frac{\rho}{2}\left\|-\sum_{i=1}^{N}\mathbf{A}_{i}\mathbf{x}_i+\mathbf{b}+\mathbf{c}\right\|^2\nonumber
\\
&= &\sum_{i=1}^{N}\left\|\mathbf{x}_i\right\|^2+I_{\mathcal{C}}\left(\mathbf{c}\right)+\frac{\rho}{2}\left\|-\sum_{i=1}^{N}\mathbf{A}_{i}\mathbf{x}_i+\mathbf{b}+\mathbf{c}+\frac{\boldsymbol{\lambda}}{\rho}\right\|^2-\frac{1}{2\rho}\left\|\boldsymbol{\lambda}\right\|^2,
\end{IEEEeqnarray}
where $\boldsymbol{\lambda}$ represents the Lagrangian multiplier, $\rho$ is a penalty parameter that tunes the severity of the quadratic penalty on constraint violations.

In line with the PJ-ADMM framework \cite{deng2017parallel,yang2022low}, the standard PJ-ADMM iterations that minimize the augmented Lagrangian function of the PM-SLP problem for QAM modulation are
\begin{IEEEeqnarray}{rCl}
\label{eq_PJADMM}
\IEEEyesnumber\IEEEyessubnumber*
\mathbf{c}^{t+1}&=&\arg \min_{\mathbf{c}}\mathcal{L}_\rho\left(\mathbf{x}^{t}_1,\cdots,\mathbf{x}^{t}_N,\mathbf{c},\boldsymbol{\lambda}^{t}\right),
\\
\mathbf{x}^{t+1}_i&=&\arg \min_{\mathbf{x}_i}\mathcal{L}_\rho\left(\mathbf{x}^{t}_{\neq i},\mathbf{x}_i,\mathbf{c}^{t+1},\boldsymbol{\lambda}^{t}\right)+\frac{1}{2}\left\|\mathbf{x}_i-\mathbf{x}^t_i\right\|^2_{\mathbf{P}_i},\forall i,
\\
\label{eq_lambda_updateQAM}
\boldsymbol{\lambda}^{t+1}&=&\boldsymbol{\lambda}^{t}+\beta\rho\left(-\sum_{i=1}^{N}\mathbf{A}_{i}\mathbf{x}^{t+1}_i+\mathbf{b}+\mathbf{c}^{t+1}\right),
\end{IEEEeqnarray}
where $\beta>0$ is a damping parameter, $\mathbf{P}_i$ is a symmetric and positive semi-definite matrix that determines the proximity between two iterations of the transmit signal and $\left\|\mathbf{x}_i\right\|^2_{\mathbf{P}_i}\triangleq\mathbf{x}^T_i\mathbf{P}_i\mathbf{x}_i$. 

Based on the above derivations, the original PM-SLP problem for QAM modulation is decomposed into multiple subproblems that can be calculated in a parallel and distributed manner with (\ref{eq_PJADMM}). In what follows, we shall derive closed-form solutions for each subproblem in the standard PJ-ADMM iterations.

The update for the slack variable $\mathbf{c}$ can be written as
\begin{IEEEeqnarray}{rCl}
\mathbf{c}^{t+1}=\arg \min_{\mathbf{c}\in\mathcal{C}}\frac{\rho}{2}\left\|-\sum_{i=1}^{N}\mathbf{A}_{i}\mathbf{x}^{t}_i+\mathbf{b}+\mathbf{c}+\frac{\boldsymbol{\lambda}^{t}}{\rho}\right\|^2,
\end{IEEEeqnarray}
which is equivalent to projecting the vector $\sum_{i=1}^{N}\mathbf{A}_{i}\mathbf{x}^{t}_i-\mathbf{b}-\frac{\boldsymbol{\lambda}^{t}}{\rho}$ onto $\mathcal{C}$, denoted by $\Pi_{\mathcal{C}}\left(\sum_{i=1}^{N}\mathbf{A}_{i}\mathbf{x}^{t}_i-\mathbf{b}-\frac{\boldsymbol{\lambda}^{t}}{\rho}\right)$. Its closed-form solution is given by
\begin{IEEEeqnarray}{rCl}
\label{eq_c_updateQAM}
\mathbf{c}^{t+1}_j=
\begin{cases}
\max\left\{\sum_{i=1}^{N}\bar{\mathbf{A}}^j_{i}\mathbf{x}^{t}_i-b_j-\frac{\lambda^{t}_j}{\rho},0\right\},&\text{if } j\in\mathcal{W},\\
0, &\text{otherwise},
\end{cases}
\IEEEeqnarraynumspace
\end{IEEEeqnarray}
where $\bar{\mathbf{A}}^j_{i}$ represents the $j$-th row of $\mathbf{A}_{i}$. $b_j$ and $\lambda^{t}_j$ are the $j$-th entries of $b_j$ and $\lambda^{t}_j$, respectively.

The iteration for $\mathbf{x}^{t+1}_i$ is updated as follows:
\begin{IEEEeqnarray}{rCl}
\mathbf{x}^{t+1}_i&=&\arg \min_{\mathbf{x}_i} {\left\|\mathbf{x}_i\right\|}^2+\frac{\rho}{2}\left\|-\mathbf{A}_{i}\mathbf{x}_i-\sum_{j\neq i}^{N}\mathbf{A}_{j}\mathbf{x}^{t}_j+\mathbf{b}+\mathbf{c}^{t+1}+\frac{\boldsymbol{\lambda}^{t}}{\rho}\right\|^2+\frac{1}{2}\left\|\mathbf{x}_i-\mathbf{x}^t_i\right\|^2_{\mathbf{P}_i},\forall i,
\IEEEeqnarraynumspace
\end{IEEEeqnarray}
which admits a closed-form solution by setting the gradient of the objective function with respect to $\mathbf{x}_i$ to zero, i.e.,
\begin{IEEEeqnarray}{rCl}
2\mathbf{x}_i+\rho\mathbf{A}^T_{i}\left(\mathbf{A}_{i}\mathbf{x}_i+\sum_{j\neq i}^{N}\mathbf{A}_{j}\mathbf{x}^{t}_j-\mathbf{b}-\mathbf{c}^{t+1}-\frac{\boldsymbol{\lambda}^{t}}{\rho}\right)+\mathbf{P}_i\left(\mathbf{x}_i-\mathbf{x}^t_i\right)=0,\forall i.
\end{IEEEeqnarray}
After some calculations, the closed-form solution for $\mathbf{x}^{t+1}_i$ can be written as
\begin{IEEEeqnarray}{rCl}
\label{xiupdate}
\mathbf{x}^{t+1}_i&=&\left(2\mathbf{I}+\rho\mathbf{A}^T_i\mathbf{A}_i+\mathbf{P}_i\right)^{-1}\left[\mathbf{P}_i\mathbf{x}^{t}_i+\rho\mathbf{A}^T_i\left(-\sum_{j\neq i}^{N}\mathbf{A}_{j}\mathbf{x}^{t}_j+\mathbf{b}+\mathbf{c}^{t+1}+\frac{\boldsymbol{\lambda}^{t}}{\rho}\right)\right],\forall i.
\end{IEEEeqnarray}
As mentioned in \cite{yang2022low}, if we take $N=2N_t$, then the transmit signal vector $\mathbf{x}$ is decomposed into $2N_t$ scalars, thus $\mathbf{A}_i$ reduces to a column vector $\mathbf{a}_i$, and $\mathbf{P}_i$ reduces to a scalar $p_i$. The update of the transmit signal can be carried out via $2N_t$ parallel and distributed scalar operations, i.e.,
\begin{IEEEeqnarray}{rCl}
x^{t+1}_i&=&\frac{p_i x^{t}_i+\rho\mathbf{a}^T_i\left(-\sum_{j\neq i}^{2N_t}\mathbf{a}_{j}x^{t}_j+\mathbf{b}+\mathbf{c}^{t+1}+\frac{\boldsymbol{\lambda}^{t}}{\rho}\right)}{2+\rho\mathbf{a}^T_i\mathbf{a}_i+p_i},\forall i.
\end{IEEEeqnarray}
Another special case is to group the real and imaginary parts of the same antenna's transmit signal into one block, then the transmit signal vector will be decomposed into $N_t$ blocks. $\mathbf{A}_i$ reduces to a $2K\times2$ matrix with orthogonal columns, which implies that the corresponding $\mathbf{A}^T_i\mathbf{A}_i$ is a $2\times 2$ diagonal matrix with equal non-zero elements. Consequently, if $\mathbf{P}_i$ is also a diagonal matrix, then the matrix inverse operation employed to update $\mathbf{x}_i$ is equivalent to taking the reciprocals of the two entries in the main diagonal, given by
\begin{IEEEeqnarray}{lCr}
\mathbf{x}^{t+1}_i=\left[\mathbf{P}_i\mathbf{x}^{t}_i+\rho\mathbf{A}^T_i\left(-\sum_{j\neq i}^{N_t}\mathbf{A}_{j}\mathbf{x}^{t}_j+\mathbf{b}+\mathbf{c}^{t+1}+\frac{\boldsymbol{\lambda}^{t}}{\rho}\right)\right]\oslash \mathbf{W},\forall i,
\end{IEEEeqnarray}
where $\mathbf{W}\triangleq diag\left(2\mathbf{I}+\rho\mathbf{A}^T_i\mathbf{A}_i+\mathbf{P}_i\right)$.

To further reduce the complexity by circumventing matrix inversion of arbitrary block size as done in \cite{yang2022low}, we set the $\mathbf{P}_i$ as $\mathbf{P}_i=\tau_i\mathbf{I}-\rho\mathbf{A}^T_i\mathbf{A}_i\mathbf{x}_i$. Accordingly, the parallel inverse-free update of $\mathbf{x}_i$ is given by
\begin{IEEEeqnarray}{rCl}
\mathbf{x}^{t+1}_i=\frac{1}{2+\tau_i}
\left[\tau_i\mathbf{x}^{t}_i+\rho\mathbf{A}^T_i\left(-\sum^{N}_{i=1}\mathbf{A}_{i}\mathbf{x}^{t}_i+\mathbf{b}+\mathbf{c}^{t+1}+\frac{\boldsymbol{\lambda}^{t}}{\rho}\right)\right], \forall i.
\IEEEeqnarraynumspace
\label{eq_primalUpdateQAM}
\end{IEEEeqnarray}

Consequently, we arrive at a PIF-SLP algorithm for the PM-SLP problem for QAM modulation.

\subsection{Duality between the PM-SLP and SB-SLP for QAM Modulation}
In this subsection, we present the duality between the PM-SLP and SB-SLP problems for QAM modulation. To begin with, the real-valued equivalent of the SB-SLP problem for QAM modulation (\ref{eq_originalSBQAM}) can be rearranged as
\begin{IEEEeqnarray}{rCl}
\label{eq_separableSBQAM}
\begin{IEEEeqnarraybox}[][c]{rCl}
\max_{\mathbf{x}_i,\mu}&\,& \quad \mu\\
\mathrm{s.t.}&\,& \sum_{i=1}^{N}\mathbf{A}_{i}\mathbf{x}_i\unrhd \mu\mathbf{b},\\
&\, &\sum_{i=1}^{N}\left\|\mathbf{x}_i\right\|^2\leq p.
\end{IEEEeqnarraybox}
\end{IEEEeqnarray}
The above formulation implies that the SB-SLP problem for QAM modulation is not separable. Recall that we proved an explicit duality between the PM-SLP and SB-SLP problems for PSK modulation in Section \ref{secDuality} and further proposed a closed-form power scaling algorithm for the SB-SLP problem for PSK modulation. Next, we shall elaborate on the same duality for QAM modulation.

Let $\mathbf{x}^{PM}$ and $p^{PM}\triangleq\|\mathbf{x}^{PM}\|^2$ denote the optimal variable and object of the PM-SLP problem for QAM modulation (\ref{eq_separablePMQAM}). $\mathbf{x}^{SB}$ and $\mu^{SB}\triangleq \min\limits_{i}\frac{1}{\bar{b}_i}\bar{\mathbf{a}}^T_i \mathbf{x}^{SB}$ are the counterparts for the SB-SLP problem for QAM modulation (\ref{eq_separableSBQAM}), where $\bar{\mathbf{a}}_i$ denotes the transpose of the $i$-th row of $\mathbf{A}$, and $\bar{b}_i$ represents the $i$-th entry of $\mathbf{b}$.
\begin{lemma}
\label{lemma_dualityvarQAM}
The PM-SLP problem (\ref{eq_separablePMQAM}) and the SB-SLP problem (\ref{eq_separableSBQAM}) are inverse problems:
\begin{IEEEeqnarray}{rCl}
\label{eq_PMSBrelaQAM}
\mathbf{x}^{PM}\left(\alpha \mathbf{b}\right)&=&\mathbf{x}^{SB}\left(\mathbf{b},p^{PM}\left(\alpha \mathbf{b}\right)\right),
\end{IEEEeqnarray}
with $\alpha=\mu^{SB}\left(\mathbf{b},p^{PM}\left(\alpha \mathbf{b}\right)\right)$. Reciprocally,
\begin{IEEEeqnarray}{rCl}
\label{eq_SBPMrelaQAM}
\mathbf{x}^{SB}\left(\mathbf{b},p\right)&=&\mathbf{x}^{PM}\left(\mu^{SB}\left(\mathbf{b},p\right)\mathbf{b}\right),
\end{IEEEeqnarray}
with $p=p^{PM}\left(\mu^{SB}\left(\mathbf{b},p\right)\mathbf{b}\right)$.
\end{lemma}
\begin{IEEEproof}[\bf Proof]
Verbatim to the proof of Lemma \ref{lemma_dualityvar}.
\end{IEEEproof}

\begin{lemma}
\label{lemma_scalQAM}
Consider the PM-SLP problem (\ref{eq_separablePMQAM}) for QAM modulation, for any $\alpha>0$, we have
\begin{IEEEeqnarray}{rCl}
\label{eq_PMvarQAM}
\mathbf{x}^{PM}\left(\alpha \mathbf{b}\right)&=&\alpha\mathbf{x}^{PM}\left(\mathbf{b}\right),\\
\label{eq_PMobjQAM}
p^{PM}\left(\alpha \mathbf{b}\right)&=&\alpha^2 p^{PM}\left(\mathbf{b}\right).
\end{IEEEeqnarray}
For the SB-SLP problem for QAM modulation,
\begin{IEEEeqnarray}{rCl}
\label{eq_SBvarQAM}
\mathbf{x}^{SB}\left(\mathbf{b},\alpha^2 p\right)&=&\alpha\mathbf{x}^{SB}\left(\mathbf{b},p\right),\\
\label{eq_SBobjQAM}
\mu^{SB}\left(\mathbf{b},\alpha^2 p\right)&=&\alpha \mu^{SB}\left(\mathbf{b},p\right).
\end{IEEEeqnarray}
\end{lemma}
\begin{IEEEproof}[\bf Proof]
Verbatim to the proof of Lemma \ref{lemma_scal}.
\end{IEEEproof}

\begin{theorem}[Duality for QAM Modulation]
\label{theorem_explicitQAM}
Let $\mathbf{x}^{PM}$ and $p^{PM}\triangleq\|\mathbf{x}^{PM}\|^2$ denote the optimal solution and the optimal object value of the PM-SLP problem for QAM modulation (\ref{eq_separablePMQAM}), respectively, then the counterparts of the SB-SLP problem, $\mathbf{x}^{SB}$ and $\mu^{SB}$, are determined as
\begin{IEEEeqnarray}{rCl}
\label{eq_SBPMvarQAM}
\mathbf{x}^{SB}\left(\mathbf{b},p\right)&=&\sqrt{\frac{p}{p^{PM}\left(\mathbf{b}\right)}}\mathbf{x}^{PM}\left(\mathbf{b}\right),\\
\label{eq_SBPMobjQAM}
\mu^{SB}\left(\mathbf{b},p\right)&=&\sqrt{\frac{p}{p^{PM}\left(\mathbf{b}\right)}}.
\end{IEEEeqnarray}
and vice versa as
\begin{IEEEeqnarray}{rCl}
\label{eq_PMSBvarQAM}
\mathbf{x}^{PM}\left(\mathbf{b}\right)&=&\frac{1}{\mu^{SB}\left(\mathbf{b},p\right)}\mathbf{x}^{SB}\left(\mathbf{b},p\right),\\
\label{eq_PMSBobjQAM}
p^{PM}\left(\mathbf{b}\right)&=&\frac{p}{\left(\mu^{SB}\left(\mathbf{b},p\right)\right)^2}.
\end{IEEEeqnarray}
\end{theorem}
\begin{IEEEproof}[\bf Proof]
Verbatim to the proof of Theorem \ref{theorem_explicit}.
\end{IEEEproof}

\begin{corollary}
\label{corollary_powscalQAM}
Similar to the PSK modulation case, the SB-SLP and PM-SLP problems for QAM modulation can also be solved simultaneously. In particular, the optimal solution to the SB-SLP problem (\ref{eq_separableSBQAM}) can be obtained by first solving the PM-SLP problem (\ref{eq_separablePMQAM}) and then scaling the transmit power to satisfy the power budget of the SB-SLP problem, and vice versa.
\end{corollary}

According to Corollary \ref{corollary_powscalQAM}, it is also feasible to solve the SB-SLP problem for QAM modulation via the closed-form power scaling algorithm, provided that the solution to the PM-SLP problem is given. In the previous section, we developed the PIF-SLP algorithm by taking advantage of the separable structure of the PM-SLP problem for QAM modulation, which can be connected with the power scaling algorithm to solve the SB-SLP problem. Therefore, we arrive at the SPIF-SLP algorithm for QAM modulation.

\section{Computational Complexity}
\label{secComplexity}
The computational overhead of the proposed PIF-SLP algorithm for QAM modulation and the closed-form power scaling algorithm is assessed by accounting for the required float-point operations, i.e., flops. The PIF-SLP algorithm for the PM-SLP problem for QAM modulation updates three variables alternately, namely, the first step updates the slack variable $\mathbf{c}$, the second step updates $N$ blocks of the transmit signal $\{\mathbf{x}_i\}$ in parallel and the last step updates the Lagrangian multiplier $\boldsymbol{\lambda}$. To simplify the analysis, it is assumed that each block of transmit signal has the same dimension, which means that the transmit signal can be decomposed into $N$ subvectors, each with $2N_t/N$ elements. Define the flop-count operator $\mathcal{F}\left(\mathbf{z}|\mathbf{y}\right)$
as the number of flops to compute $\mathbf{z}$ given $\mathbf{y}$. As a result, the proposed PIF-SLP algorithm for QAM modulation costs
\begin{IEEEeqnarray}{rCl}
\mathcal{F}\left(\boldsymbol{\lambda}^{t+1}|\boldsymbol{\lambda}^t\right)&=&\mathcal{F}\left(\mathbf{c}^{t+1}|\left\{\mathbf{A}_i\mathbf{x}^{t}_i,\boldsymbol{\lambda}^{t}\right\}\right)+\mathcal{F}\left(\mathbf{A}_i\mathbf{x}^{t+1}_i|\left\{\mathbf{A}_i\mathbf{x}^{t}_i,\mathbf{c}^{t+1},\boldsymbol{\lambda}^t\right\}\right)+\mathcal{F}\left(\boldsymbol{\lambda}^{t+1}|\left\{\boldsymbol{\lambda}^t,\mathbf{A}_i\mathbf{x}^{t+1}_i,\mathbf{c}^{t+1}\right\}\right)\nonumber\\
&=&\mathcal{O}(2K)+\mathcal{O}((2K+1)2N_t/N)+\mathcal{O}(2K)
\end{IEEEeqnarray}
flops per iteration, which is the same as the PIF-SLP algorithm for PSK signaling proposed in \cite{yang2022low}. As for the closed-form power scaling algorithm for the SB-SLP problem, (\ref{eq_SBPMvar}) requires $\mathcal{O}(2N_t)$ flops.

\section{Simulation Results}
\label{secResults}
Numerical simulations are conducted in this section to evaluate the performance of the proposed SPIF-SLP algorithm for SB-SLP problem for PSK modulation, along with the proposed PIF-SLP algorithm and the SPIF-SLP algorithm for PM-SLP and SB-SLP problems for QAM modulation, respectively. Without loss of generality, QPSK and 16QAM are selected as representative schemes for PSK and QAM modulations, respectively. The i.i.d. data symbols in $\tilde{\mathbf{s}}$ are drawn from the normalized QPSK constellation, i.e., $\mathcal{M}=4$ or 16QAM constellation, i.e., $\mathcal{M}=16$. We use `$K\times N_t$' to denote a downlink system with $K$ single-antenna users and an $N_t$-antenna BS. For both PIF-SLP and SPIF-SLP in all the considered scenarios, we choose $\tau_i=\tau=0.8\rho\left\|\mathbf{A}\right\|^2, \forall i$. The damping parameter $\beta$ is set to 1. Unless otherwise specified, the penalty parameter $\rho$ is set to 0.3, 0.4, 0.06, 0.03, and 0.015 for $8\times8$, $12\times12$, $12\times16$, $24\times32$, and $48\times64$ MIMO configurations, respectively. For the SB-SLP problem simulations, the square roots of the weights $\frac{1}{\sqrt{\gamma_k}}$ are all set to 1. We assume each random channel realization is used to transmit one frame of data symbols, where each frame contains $N_s=20$ symbol slots \cite{3gpp2017study}. Note that it is beyond the scope of this work to implement the proposed parallelizable algorithms in physical parallel processing units. However, if we use the \emph{for} or \emph{parfor} loop in  MATLAB to simulate the low-complexity parallel procedure for the considered scenarios, the loop itself will cost a big portion of time due to sequential implementation or parallelization overheads. Therefore we set $N=1$ in this section.

\begin{figure*}[!t]
\begin{minipage}[t]{0.3\linewidth}
\vspace{0pt}
\centering
\setlength{\abovecaptionskip}{-1cm}
\includegraphics[width=\textwidth]{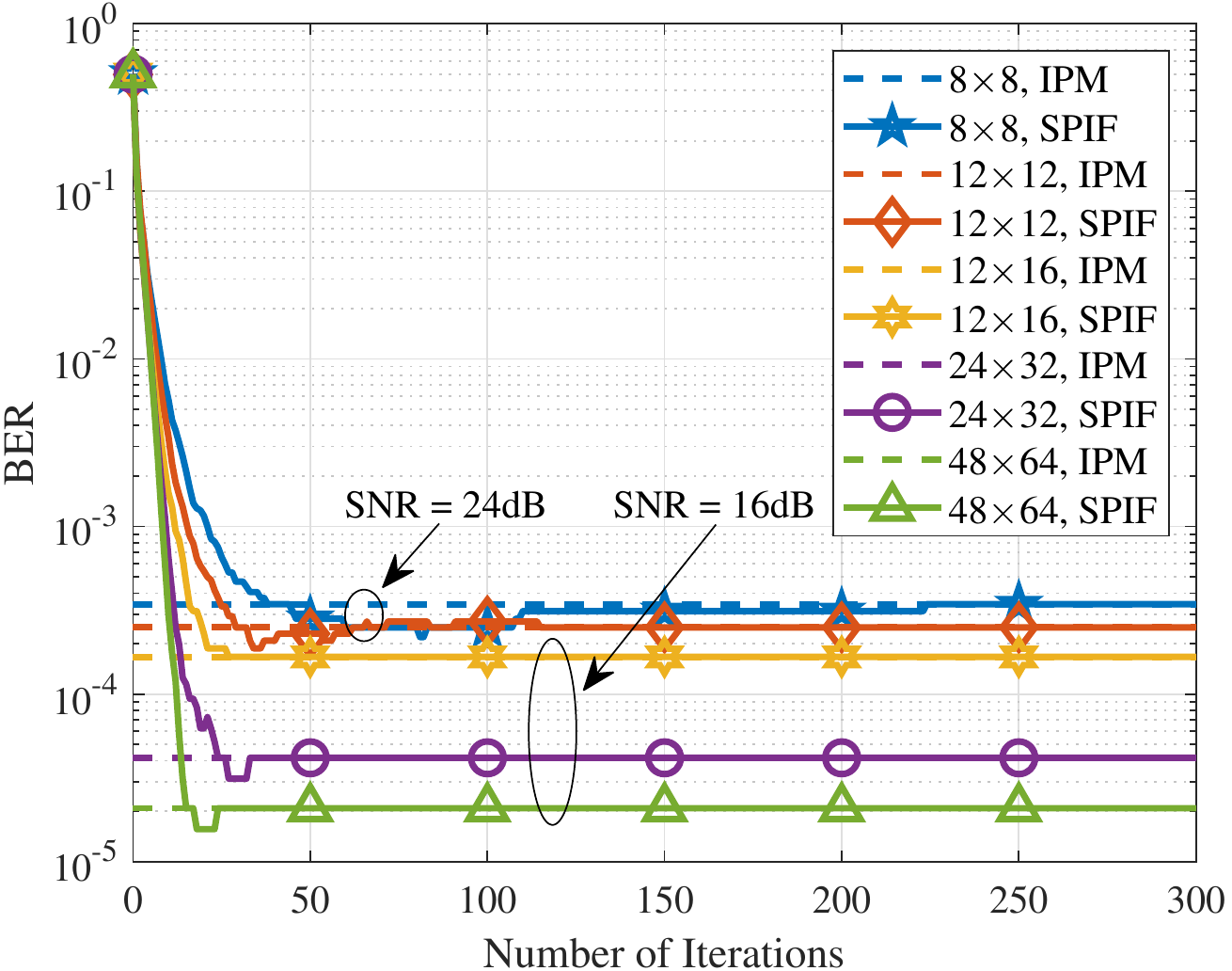}
\caption{BER versus number of iterations, $\mathrm{SNR}=24$dB for fully-loaded systems, $\mathrm{SNR}=16$dB for under-loaded systems, $N_c=100$, $N_s=20$, QPSK.}
\label{fig_BERiter}
\end{minipage}
\hspace{0.1cm}
\begin{minipage}[t]{0.3\linewidth}
\vspace{0pt}
\centering
\setlength{\abovecaptionskip}{-1cm}
\includegraphics[width=\textwidth]{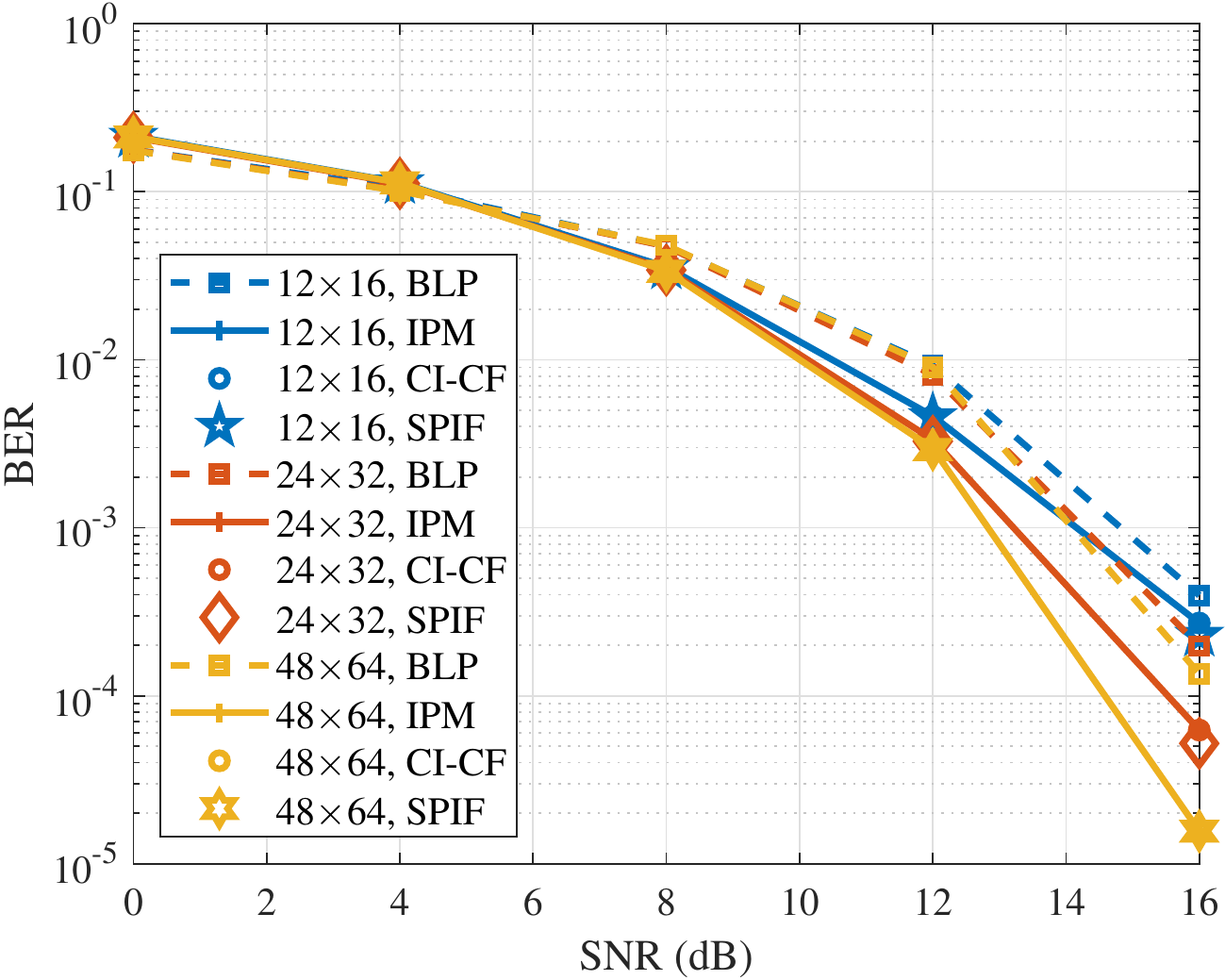}
\caption{BER versus SNR for three under-loaded MIMO configurations, $T=40$, $N_c=100$, $N_s=20$, QPSK.}
\label{fig_BER}
\end{minipage}
\hspace{0.1cm}
\begin{minipage}[t]{0.3\linewidth}
\vspace{0pt}
\centering
\setlength{\abovecaptionskip}{-1cm}
\includegraphics[width=\textwidth]{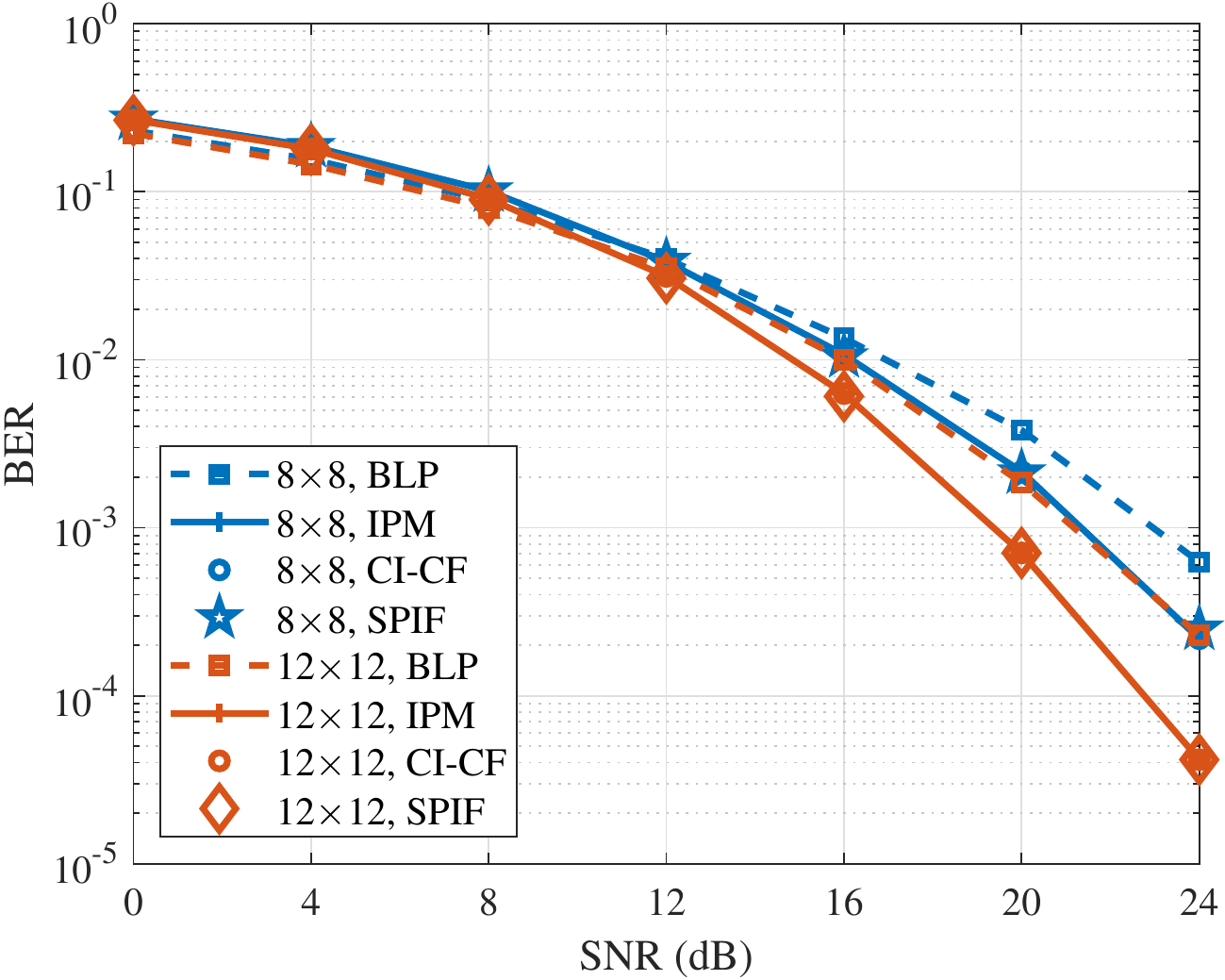}
\caption{BER versus SNR for two fully-loaded MIMO configurations, $\Delta<1\times10^{-2}$, $T_{max}=100$, $N_c=100$, $N_s=20$, QPSK.}
\label{fig_BERfull}
\end{minipage}
\end{figure*}

To demonstrate the convergence of the proposed SPIF-SLP algorithm for PSK modulation, we first study its bit error rate (BER) performance as a function of the number of iterations, the results are averaged over 2000 symbol slots, where the number of random channel realizations $N_c=100$. The benchmark scheme is the IPM implemented by the CVX software package \cite{grant2014cvx}. Fig. \ref{fig_BERiter} presents the BER results for the aforementioned five MIMO configurations. The required number of iterations for the BER of the SPIF-SLP algorithm converging to that of the IPM is about $T=40$ for the three considered under-loaded MIMO configurations. The acquired number of iterations for convergence is used in the remaining under-loaded simulations for PSK modulation. On the other hand, the fully-loaded MIMO configurations take more iterations to converge because of the symmetric channel. It is worth noting that in each fully-loaded simulation, the number of iterations for convergence is different.

We then assess the BER performance versus SNR as well as the time complexity in terms of the average execution time of the proposed SPIF-SLP algorithm for PSK modulation. The benchmark schemes are selected as the conventional linear SB-BLP solved by line search \cite{wiesel2005linear,6832894}, the SB-SLP solved by IPM \cite{grant2014cvx}, and the closed-form solutions for CI precoding (CI-CF) \cite{li2018interference}. For the SB-SLP simulations, the total power budget $p$ is set to 1 in different scenarios, and the noise variance varies depending on the SNR. 

Fig. \ref{fig_BER} depicts the BER performance versus the increasing SNR for the aforementioned three under-loaded MIMO configurations. It can be observed that the BER performance of the proposed SPIF-SLP algorithm is almost consistent with that of the selected benchmark algorithms, which validates the effectiveness of the proposed SPIF-SLP algorithm for PSK modulation in under-loaded scenarios.

As for the aforementioned two fully-loaded MIMO configurations, Fig. \ref{fig_BERfull} illustrates their BER performance as a function of the increasing SNR. The stopping criterion is set to $\Delta<1\times10^{-2}$, where $\Delta\triangleq\|\mathbf{x}^t-\mathbf{x}^{t-1}\|$ denotes the iteration decrease, and the maximum number of iterations $T_{max}=100$. The fully-loaded system may have strong interfering channels, which long for a more accurate transmit signal than the under-loaded system to exploit interference. This criterion provides an acceptable accuracy while keeping a reasonable iteration scale. We can observe in Fig. \ref{fig_BERfull} that the proposed SPIF-SLP algorithm is also competent for the fully-loaded systems with PSK signaling.

\begin{table}[!htbp] 
\renewcommand{\arraystretch}{1.3}
\caption{Average Execution Time per Frame in Sec. for SB-SLP, SNR = 24dB for fully-loaded systems, SNR = 16dB for under-loaded systems, $N_c=100$, $N_s=20$, QPSK.}
\label{table_timeSB}
\centering
\begin{tabular}{c  r@{.}l r@{.}l r@{.}l r@{.}l r@{.}l}
\toprule 
\multicolumn{1}{c}{\multirow{1}{*}{}}& \multicolumn{2}{c}{8 $\times$ 8}& \multicolumn{2}{c}{12 $\times$ 12}& \multicolumn{2}{c}{12 $\times$ 16} & \multicolumn{2}{c}{24 $\times$ 32}& \multicolumn{2}{c}{48 $\times$ 64}\\
\hline 
\multicolumn{1}{l}{BLP}&9&7143&12&6695&6&9729&19&1661&227&1547\\
\multicolumn{1}{l}{IPM}&9&8468&9&1228&9&0986&8&3034&9&7937\\   
\multicolumn{1}{l}{CI-CF}&1&3224e-2&1&8697e-1&5&2829e-2&4&0618e-2&1&0721e-1\\
\multicolumn{1}{l}{\bf SPIF}&\bf 2&3793e-3&\bf 3&1054e-3&\bf 2&7303e-3
&\bf 7&5000e-3& \bf 5&6801e-2
\\
\bottomrule
\end{tabular}
\end{table}

Table \ref{table_timeSB} lists the time complexity in terms of the average execution time per frame of the compared algorithms for the SB-SLP problem for PSK modulation under five MIMO configurations, where the number of iterations of the SPIF-SLP algorithm is the same as in Fig. \ref{fig_BER} and Fig. \ref{fig_BERfull}. The execution time of the proposed SPIF-SLP algorithm for PSK modulation is about 18.0\%, 1.7\%, 5.2\%, 18.5\%, and 53.0\% of that of the CI-CF algorithm in $8\times8$, $12\times12$, $12\times16$, $24\times32$, and $48\times64$ MIMO configurations, respectively. The complexity reduction of the proposed SPIF-SLP algorithm for PSK modulation is appealing in all the considered MIMO configurations.

Next, the average transmit power and the time complexity in terms of the average execution time per frame of the proposed PIF-SLP algorithm for QAM modulation are investigated. The results are compared with those of the conventional linear PM-BLP \cite{wiesel2005linear,6832894}, the IPM implemented by the CVX software package \cite{grant2014cvx}, and the efficient gradient projection algorithm (EGPA) \cite{masouros2015exploiting}. Consider unit noise variance along with equal instantaneous SINR threshold for each user, i.e., $\sigma^2_k=\sigma^2=1$, $\gamma_k=\gamma,\forall k$.  

\begin{figure*}[!t]
\begin{minipage}[t]{0.3\linewidth}
\vspace{0pt}
\centering
\setlength{\abovecaptionskip}{-1cm}
\includegraphics[width=\textwidth]{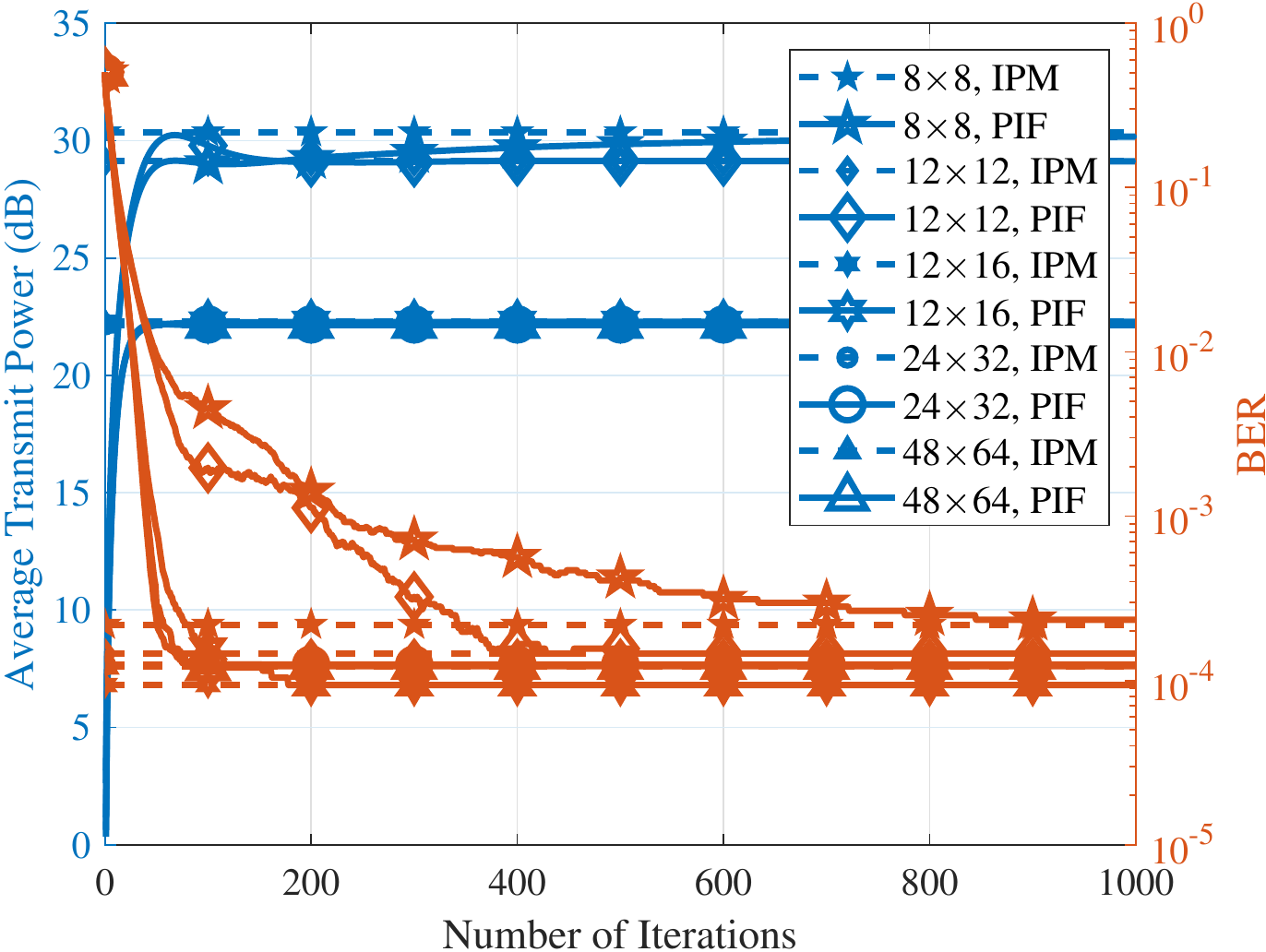}
\caption{Average transmit power and BER versus number of iterations for various MIMO configurations, $Nc=100$, $N_s=20$, 16QAM.}
\label{fig_POWBERiter}
\end{minipage}
\hspace{0.1cm}
\begin{minipage}[t]{0.3\linewidth}
\vspace{0pt}
\centering
\setlength{\abovecaptionskip}{-1cm}
\includegraphics[width=\textwidth]{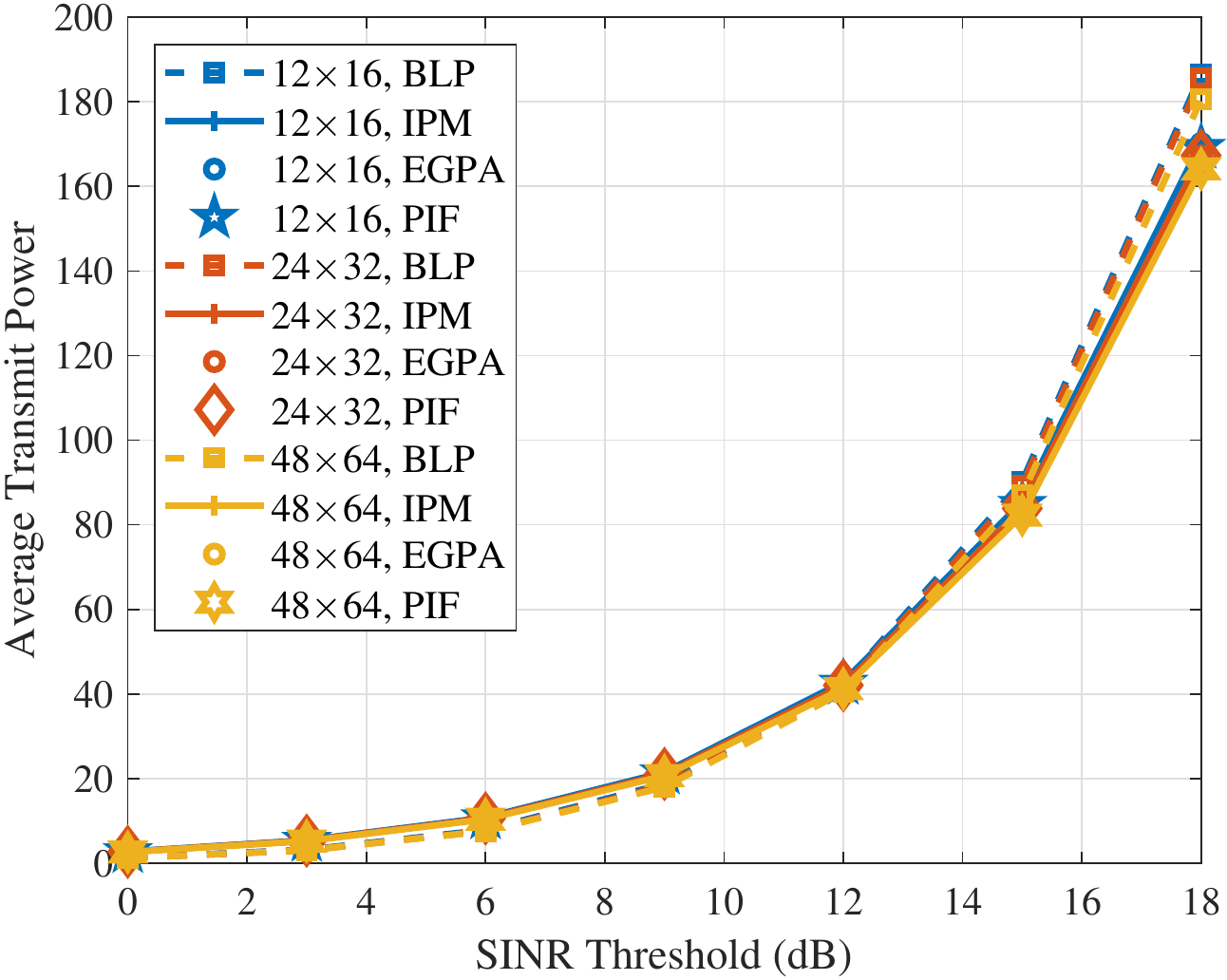}
\caption{Average transmit power versus SINR threshold for three under-loaded MIMO configurations, $T=150$, $Nc=100$, $N_s=20$, 16QAM.}
\label{fig_POW}
\end{minipage}
\hspace{0.1cm}
\begin{minipage}[t]{0.3\linewidth}
\vspace{0pt}
\centering
\setlength{\abovecaptionskip}{-1cm}
\includegraphics[width=\textwidth]{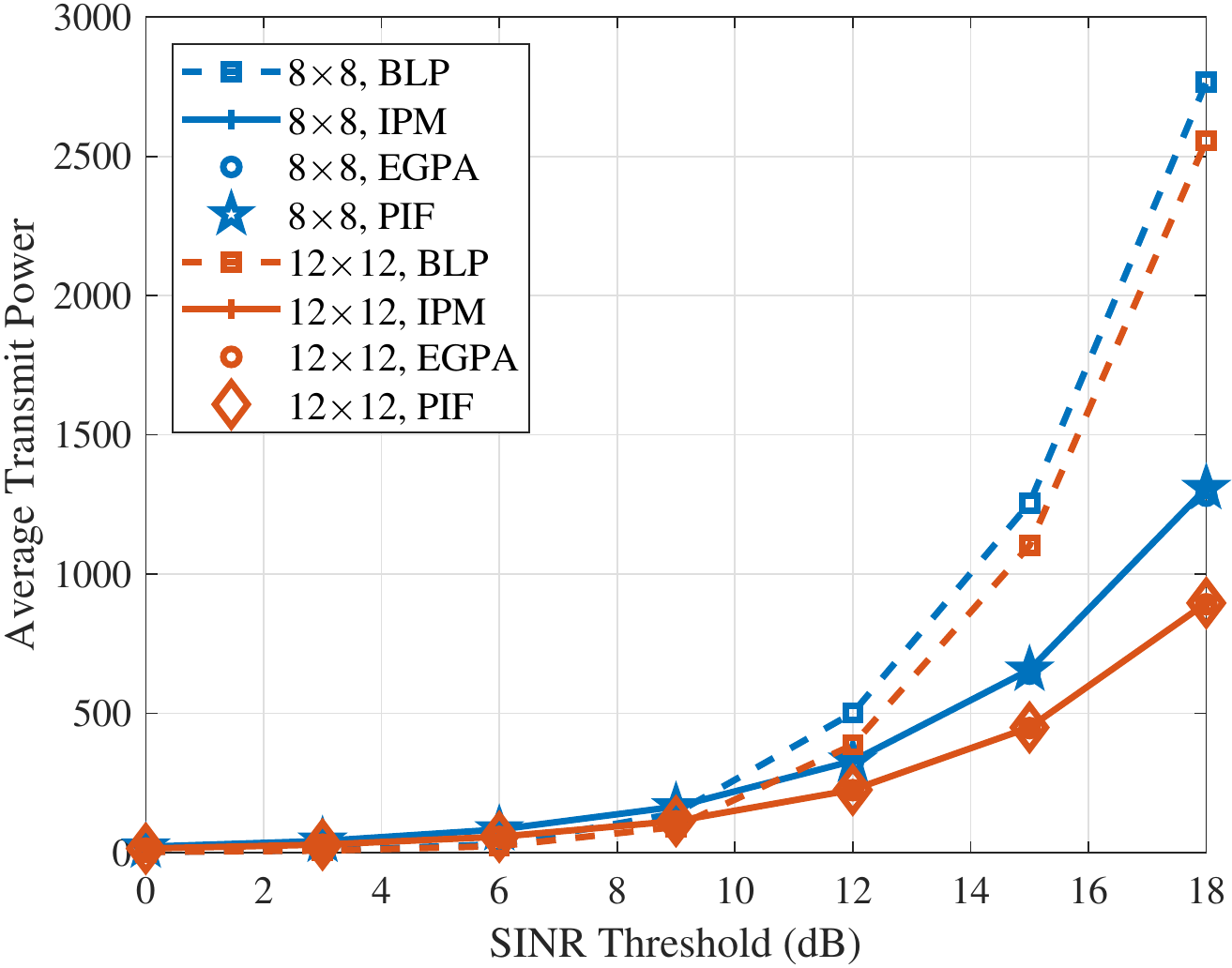}
\caption{Average transmit power versus SINR threshold for two fully-loaded MIMO configurations, $\Delta<1\times10^{-7}$ for $8\times8$, $\Delta<1\times10^{-6}$ for $12\times12$, $T_{max}=4000$, $N_c=100$, $N_s=20$, 16QAM.}
\label{fig_POWfull}
\end{minipage}
\end{figure*}

Again, to demonstrate the convergence of the proposed PIF-SLP algorithm for QAM modulation, we present the average transmit power and the BER performance versus the number of iterations for four MIMO configurations in Fig. \ref{fig_POWBERiter}, where the results are averaged over 100 random channel realizations. The SINR threshold is set to 18dB. The penalty parameter is set to 0.8 for both $8\times8$ and $12\times12$ MIMO configurations. For the two under-loaded MIMO configurations, it can be seen that the PIF-SLP algorithm takes about 150 iterations to approach the benchmark scheme. Thousands of iterations are needed by the two challenging fully-loaded MIMO configurations to converge. Compared to the previous PSK case, the proposed algorithm for QAM modulation needs more iterations to converge. Because the correct detection of QAM modulation relies on both the amplitude and phase of the received signal, thus needs a more accurate transmit signal.

Fig. \ref{fig_POW} shows the average transmit power performance of the proposed PIF-SLP algorithm for QAM modulation as a function of the SINR threshold for three under-loaded MIMO configurations. It can be observed that the performance of the proposed PIF-SLP algorithm matches that of the IPM and the EGPA at various MIMO configurations and SINR thresholds, namely, the optimality of the proposed PIF-SLP algorithm for QAM modulation is guaranteed after adequate iterations. The average transmit power of the PM-SLP problem is substantially affected by the system load. Although the three MIMO configurations have different numbers of transmit and receive antennas, they consume almost equal transmit power to serve different numbers of users owing to the same system load level.

Fig. \ref{fig_POWfull} shows the average transmit power performance of the proposed PIF-SLP algorithm for QAM modulation as a function of the SINR threshold for two fully-loaded MIMO configurations, where the penalty parameter is set to 0.8. Similar to the under-loaded case presented in Fig. \ref{fig_POW}, the performance of the proposed PIF-SLP algorithm is well-matched with the benchmark SLP schemes. The BER gap between the conventional linear BLP and SLP is more prominent in the fully-loaded case, where there is more interference to be exploited.

\begin{table}[!htbp] 
\renewcommand{\arraystretch}{1.3}
\caption{Average Execution Time in Sec. for PM-SLP, $\gamma=18$dB, $N_c=100$, $Ns=20$, 16QAM.}
\label{table_timePM}
\centering
\begin{tabular}{c  r@{.}l r@{.}l r@{.}l r@{.}l r@{.}l}
\toprule 
\multicolumn{1}{c}{\multirow{1}{*}{}}& \multicolumn{2}{c}{8 $\times$ 8}& \multicolumn{2}{c}{12 $\times$ 12}& \multicolumn{2}{c}{12 $\times$ 16}& \multicolumn{2}{c}{24 $\times$ 32} & \multicolumn{2}{c}{48 $\times$ 64}\\
\hline 
\multicolumn{1}{l}{BLP}&3&2739e-1&3&9259e-1& 4&9426e-1& 1&1478& 17&2852\\   
\multicolumn{1}{l}{IPM}&3&6611&3&4122& 4&3867& 3&5160& 5&8538\\   
\multicolumn{1}{l}{EGPA}&2&3247&3&1020& 5&2505e-1& 2&5729& 18&6100\\
\multicolumn{1}{l}{\bf PIF}&\bf 4&3854e-2
&\bf 5&3732e-2& \bf 9&5013e-3& \bf 1&3688e-2& \bf 1&5444e-1
\\
\bottomrule
\end{tabular}
\end{table}

Table \ref{table_timePM} further compares the average execution time per frame of the considered algorithms for the PM-SLP problem for QAM modulation, where the parameters are the same as those in Fig. \ref{fig_POW} and Fig. \ref{fig_POWfull}. Due to its simple and inverse-free processing, the proposed PIF-SLP algorithm has the fastest speed to solve the PM-SLP problem for QAM modulation compared to the IPM and the EGPA. In particular, the proposed PIF-SLP algorithm can provide a processing time that is 53.01, 57.73, 55.26, 187.97, and 120.50 times faster than the EGPA in $8 \times 8$, $12 \times 12$, $12 \times 16$, $24 \times 32$, and $48 \times 64$ MIMO configurations, respectively. It can also provide a processing time that is 83.48, 63.50, 461.69, 256.87, and 37.90 times faster than the IPM in $8 \times 8$, $12 \times 12$, $12 \times 16$, $24 \times 32$, and $48 \times 64$ MIMO configurations, respectively. Note that the execution time of the PIF-SLP algorithm can further be greatly reduced by parallel implementation in practice. 

\begin{figure*}[!t]
\begin{minipage}[t]{0.3\linewidth}
\vspace{0pt}
\centering
\setlength{\abovecaptionskip}{-1cm}
\includegraphics[width=\textwidth]{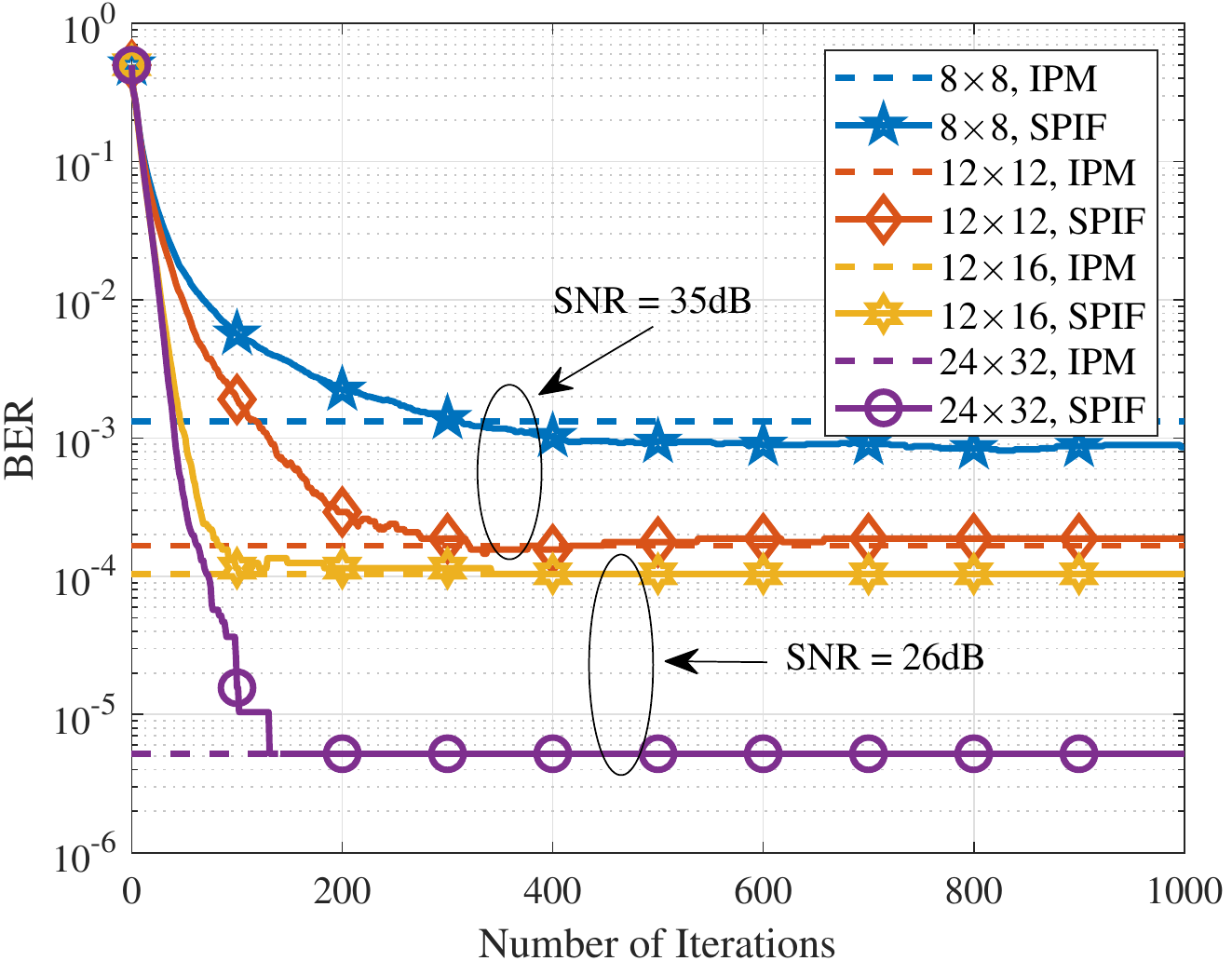}
\caption{BER versus number of iterations, $N_c=100$, $N_s=20$, 16QAM.}
\label{fig_BERiterQAM}
\end{minipage}
\hspace{0.1cm}
\begin{minipage}[t]{0.3\linewidth}
\vspace{0pt}
\centering
\setlength{\abovecaptionskip}{-1cm}
\includegraphics[width=\textwidth]{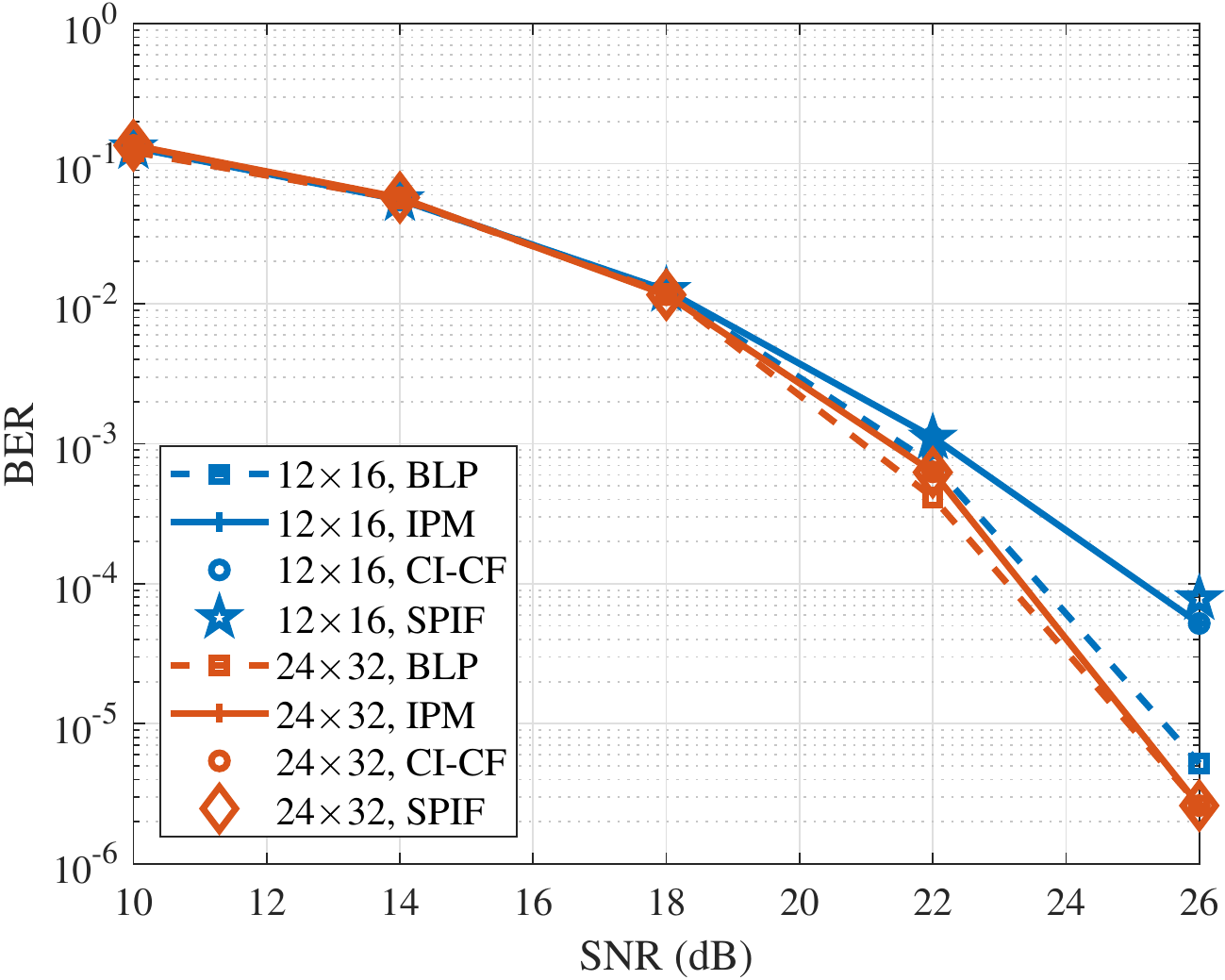}
\caption{BER versus SNR for three under-loaded MIMO configurations, $T=150$, $N_c=200$, $N_s=20$, 16QAM.}
\label{fig_BERQAM}
\end{minipage}
\hspace{0.1cm}
\begin{minipage}[t]{0.3\linewidth}
\vspace{0pt}
\centering
\setlength{\abovecaptionskip}{-1cm}
\includegraphics[width=\textwidth]{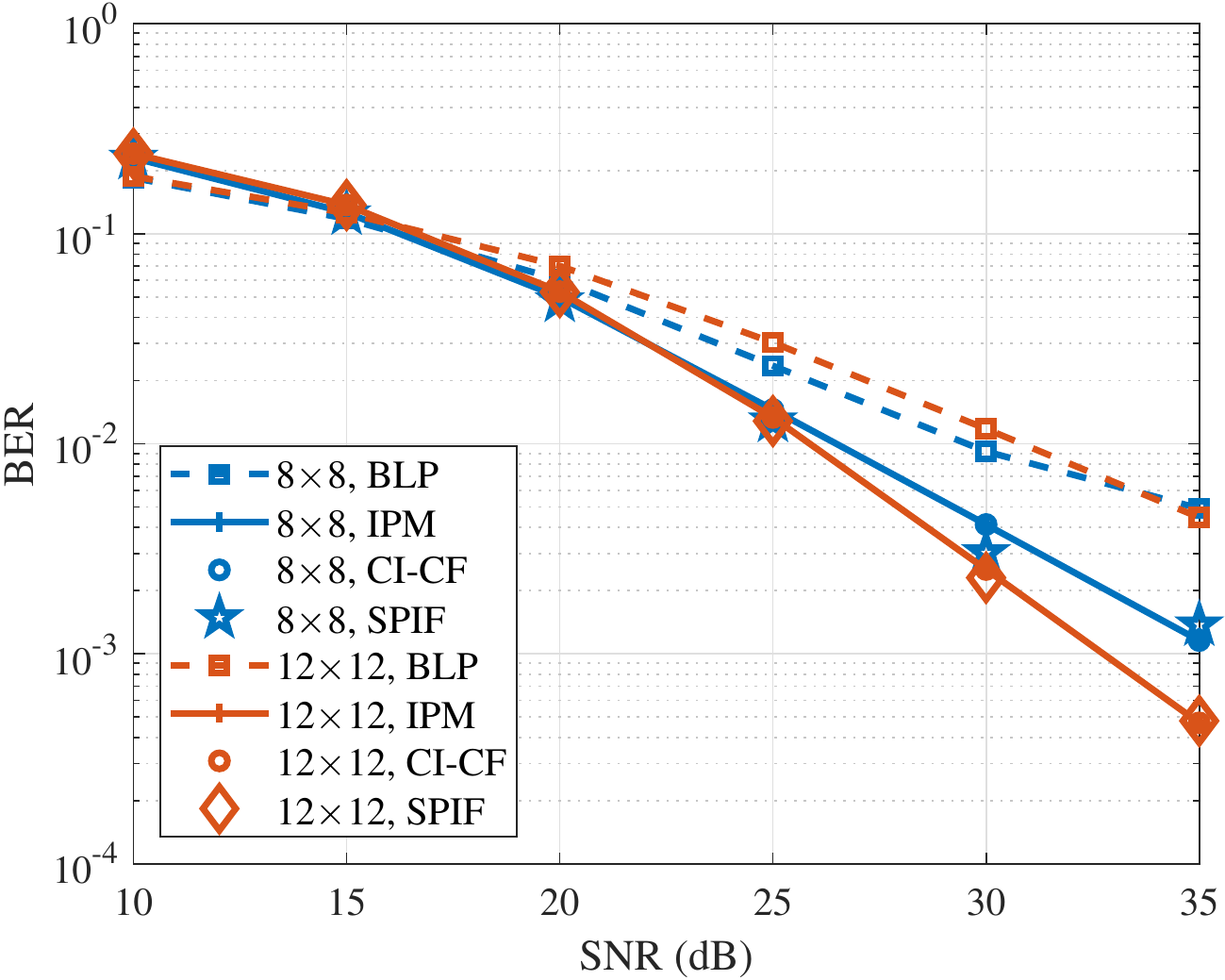}
\caption{BER versus SNR for two fully-loaded MIMO configurations, $\Delta<1\times10^{-4}$ for $8\times8$, $\Delta<1\times10^{-3}$ for $12\times12$, $T_{max}=300$, $N_c=100$, $N_s=20$,  16QAM.}
\label{fig_BERQAMfull}
\end{minipage}
\end{figure*}

Similar to the PSK case, for the SPIF-SLP algorithm for QAM modulation, we first evaluate the required number of iterations to converge in Fig. \ref{fig_BERiterQAM}, where the BER is depicted as a function of the number of iterations. The results are averaged over 2000 symbol slots. The benchmark scheme is the IPM implemented by the CVX software package \cite{grant2014cvx}. We set $\mathrm{SNR}=18$dB and $\mathrm{SNR}=35$dB for under-loaded and fully-loaded MIMO configurations, respectively. After 150 iterations, the BER of the SPIF-SLP algorithm for three under-loaded MIMO configurations can converge to that of the IPM. Therefore, we set the number of iterations of the SPIF-SLP algorithm to 150 for the remaining under-loaded simulations. The fully-loaded MIMO configurations take far more iterations to converge than the under-loaded systems.

Fig. \ref{fig_BERQAM} depicts the BER performance of the proposed SPIF-SLP algorithm for QAM modulation versus the increasing SNR for the aforementioned three under-loaded MIMO configurations. The benchmark schemes are selected as the same as the PSK case. The same trends can be seen in Fig. \ref{fig_BERQAM} and Fig. \ref{fig_BER}, the BER performance of the proposed SPIF-SLP algorithm is almost consistent with that of the selected benchmark SLP algorithms, which validates the effectiveness of the proposed SPIF-SLP algorithm for QAM modulation. We can see that the BLP has lower BER than SLP in Fig. \ref{fig_BERQAM}. This can be explained by two reasons. First, we impose average power and symbol-level power constraints on BLP and SLP, respectively. Second, the CI gain is less significant in QAM than in PSK, because fewer constellation points can achieve CI.

Fig. \ref{fig_BERQAMfull} depicts the BER performance of the proposed SPIF-SLP algorithm for QAM modulation versus the increasing SNR for two fully-loaded MIMO configurations. It can be observed that the SPIF algorithm can approach the performance of the IPM in fully-loaded systems with QAM signaling. The enhanced superiority of SLP over BLP can also be observed in the fully-loaded systems.

\begin{table}[!htbp] 
\renewcommand{\arraystretch}{1.3}
\caption{Average Execution Time per Frame in Sec. for SB-SLP, SNR = 35dB and $N_c=100$ for fully-loaded systems, SNR = 26dB and $N_c=200$ for under-loaded systems, $N_s=20$, 16QAM.}
\label{table_timeSBQAM}
\centering
\begin{tabular}{c  r@{.}l r@{.}l r@{.}l r@{.}l}
\toprule 
\multicolumn{1}{c}{\multirow{1}{*}{}}& \multicolumn{2}{c}{8 $\times$ 8}& \multicolumn{2}{c}{12 $\times$ 12}& \multicolumn{2}{c}{12 $\times$ 16} & \multicolumn{2}{c}{24 $\times$ 32}\\
\hline 
\multicolumn{1}{l}{BLP}& 7&7643 & 11&6703 & 5&6993 & 17&1852\\   
\multicolumn{1}{l}{IPM}& 6&8392 & 7&0782 & 4&9532 & 5&2007\\   
\multicolumn{1}{l}{CI-CF}& 1&5468e-2 & 2&3632e-2 & 1&7495e-2 & 6&2024e-2\\
\multicolumn{1}{l}{\bf SPIF}&\bf 1&6135e-2
&\bf 1&9395e-2& \bf 8&5678e-3& \bf 1&4329e-2
\\
\bottomrule
\end{tabular}
\end{table}

Table \ref{table_timeSBQAM} lists the time complexity in terms of the average execution time per frame of the compared algorithms for the SB-SLP problem for QAM modulation under four MIMO configurations, where the parameters are the same as those in Fig. \ref{fig_BERQAM} and Fig. \ref{fig_BERQAMfull}. The execution time of the proposed SPIF-SLP algorithm for QAM modulation is about 104.3\%, 82.1\%, 49.0\%, and 23.1\%
 of that of the CI-CF algorithm in $8\times 8$, $12\times 12$, $12\times16$, and $24\times32$ MIMO configurations, respectively. The complexity reduction of the proposed SPIF-SLP algorithm is prominent for QAM modulation in the under-loaded MIMO configurations. For $8\times8$ and $12\times12$ MIMO configurations, the comparable execution time of the SPIF and CI-CF can be observed in Table \ref{table_timeSBQAM}. On the one hand, in the fully-loaded systems with QAM signaling, the CI-CF takes fewer iterations to converge due to the smaller search space, i.e., fewer constellation points can exploit interference compared to the PSK case. On the other hand, for the SPIF-SLP algorithm, the number of iterations of the SPIF-SLP algorithm is boosted by both the symmetric channel and QAM signaling. This can be alleviated by parallel implementation in practice.

\section{Conclusion}
\label{secConclusion}
In this work, we propose low-complexity algorithms for CI-based SLP in both PSK and QAM modulations via the PJ-ADMM framework, and present an explicit duality between two typical CI precoding problems, i.e., the PM problem as well as the max-min SB problem. The revealed separability of the PM-SLP problem for QAM modulation induces the PIF-SLP algorithm, which decomposes the problem into multiple parallel subproblems with simple closed-form solutions that are free of matrix inversion. To take advantage of this algorithm to solve the SB-SLP problem, we investigate the one-to-one mapping between the PM-SLP and SB-SLP problems. Moreover, a closed-form power scaling algorithm is developed to solve the SB-SLP problem on the prior knowledge of the corresponding PM-SLP solution. Jointly considering the power scaling algorithm and the PIF-SLP algorithm, we develop a SPIF-SLP algorithm for the SB-SLP problem. Through numerical results, both the PIF-SLP and SPIF-SLP algorithms are shown to provide low-complexity features while guaranteeing optimal performance.

\ifCLASSOPTIONcaptionsoff
  \newpage
\fi

\bibliographystyle{IEEEtran}
\bibliography{IEEEabrv,references}

\end{document}